\documentclass[11pt]{article}
\usepackage{latexsym,amssymb,amstext,amsmath}
\usepackage{hyperref}
\renewcommand{\baselinestretch}{1.2}
\allowdisplaybreaks
\def\Slash#1{\rlap{\hbox{$\mskip 3 mu /$}}#1}      
\newcommand{\ft}[2]{{\textstyle\frac{#1}{#2}}}

\begin{document}
%
\begin{titlepage}
\begin{flushright} \small
 ITP-UU-11/44 \\  Nikhef-2011-031\\CPHT-RR106.1211 
\end{flushright}
\bigskip

\begin{center}
 {\LARGE\bfseries  The off-shell 4D/5D connection}
\\[10mm]
\textbf{Nabamita Banerjee$^a$, Bernard de Wit$^{a,b}$ and Stefanos
  Katmadas$^{c}$ }\\[5mm] 
\vskip 4mm
$^a${\em Institute for Theoretical Physics, Utrecht
  University,} \\
  {\em Leuvenlaan 4, 3584 CE Utrecht, The Netherlands}\\
$^b${\em Nikhef Theory Group, Science Park 105, 1098 XG Amsterdam, The
  Netherlands}\\
$^c${\em  Centre de Physique Th\'eorique, Ecole Polytechnique, CNRS,}\\
{\em 91128 Palaiseau, France}\\[3mm]
{\tt N.Banerjee@uu.nl}\,,\;{\tt B.deWit@uu.nl}\,,\; 
{\tt katmadas@cpht.polytechnique.fr} 
\end{center}

\vspace{3ex}

\begin{center}
{\bfseries Abstract}
\end{center}
\begin{quotation} \noindent A systematic off-shell reduction scheme
  from five to four space-time dimensions is presented for
  supergravity theories with eight supercharges. It is applicable to
  theories with higher-derivative couplings and it is used to address
  a number of open questions regarding BPS black holes in five
  dimensions. Under this reduction the $5D$ Weyl multiplet becomes
  reducible and decomposes into the $4D$ Weyl multiplet and an extra
  Kaluza-Klein vector multiplet. The emergence of the pseudoscalar
  field of the latter multiplet and the emergence of the $4D$
  R-symmetry group are subtle features of the reduction. The reduction
  scheme enables to determine how a $5D$ supersymmetric Lagrangian
  with higher-derivative couplings decomposes upon dimensional
  reduction into a variety of independent $4D$ supersymmetric
  invariants, without the need for imposing field equations. In this
  way we establish, for example, the existence of a new N=2
  supersymmetric invariant that involves the square of the Ricci
  tensor. Finally we resolve the questions associated with the $5D$
  Chern-Simons terms for spinning BPS black holes and their relation to
  the corresponding $4D$ black holes.
\end{quotation}

\vfill

\end{titlepage}

\section{Introduction}
\label{sec:introduction}
\setcounter{equation}{0}
Dimensional reduction plays an important role in the study of many
aspects of supergravity and string theory. Usually the reduction is
performed in the context of supersymmetric {\it on-shell} field
configurations. For theories with a large number of supercharges this
is unavoidable, as off-shell representations are usually not
available. For theories based on off-shell representations there has
been little effort so far to define a suitable dimensional reduction
scheme, because the extra auxiliary fields contained in the off-shell
configuration can be ignored upon solving their corresponding
(algebraic) field equations. However, in the presence of
higher-derivative couplings, these field equations are no longer
algebraic. In their on-shell form these couplings will therefore take
the form of an iterative expansion in increasing powers of space-time
derivatives, which obscures the structure of the underlying off-shell
invariants.

Dimensional reduction of {\it off-shell} configurations is based on a
corresponding reduction of the off-shell supersymmetry algebra. It can
therefore be performed systematically on separate multiplets. To
accomplish this one maps a supermultiplet in higher dimension on a
corresponding, not necessarily irreducible, supermultiplet in lower
dimension, possibly in a certain supergravity background. When
considering the supersymmetry algebra in the context of a
lower-dimensional space-time, the dimension of the automorphism group
of the algebra (the R-symmetry group) usually increases, and this has
to be taken into account when casting the resulting supermultiplet in
its standard form. The fact that irreducible multiplets in higher
dimension can become reducible in lower dimensions, further
complicates the reduction procedure.

In this paper we study the off-shell dimensional reduction of $5D$
$N=1$ superconformal multiplets to the corresponding $N=2$
superconformal multiplets in four dimensions. On-shell dimensional
reduction of these theories has been considered in the past for a
variety of reasons, but mainly in the context of actions that are at
most quadratic in space-time derivatives. For some earlier work we
refer to
\cite{deWit:1992wf,hep-th/0503217,Gaiotto:2005xt,Behrndt:2005he}. We
choose the superconformal setting because this enables us to work in
the context of the superconformal multiplet calculus, which has been
developed in both four and five dimensions.\footnote{
  In \cite{Kugo:2000hn}, off-shell dimensional reduction in $6D$ was
  used to determine the superconformal transformations in $5D$.
} 
It may seem that conformal invariance will be an impediment to
dimensional reduction, because conformal symmetry depends strongly on
the actual space-time dimension. However, it turns out that this is
not problematic at all, because the conformal transformations are
formulated in a way that is independent on the four- or
five-dimensional space-time manifold (which is subject to general
diffeomorphisms) whereas the scale transformations and conformal
boosts are defined in the tangent space. We will not elaborate on this
aspect in further detail as it will be rather explicit in the
construction that we are about to present. The construction is
somewhat facilitated by the fact that the spinor dimension is the same
in five and in four dimensions: in four dimensions we are dealing with
a doublet of four-component independent Majorana spinors, and in five
dimensions we have a four-component spinor, which can be treated
either as a Dirac spinor or as a symplectic Majorana spinor. Both
these spinors share a common $\mathrm{SU}(2)$ factor in the R-symmetry
group. We will exhibit in detail how the additional $\mathrm{U}(1)$
factor will emerge in four dimensions. Here we recall that in
conformal supergravity, R-symmetry is realized as a local symmetry.

The whole reduction scheme is subtle, especially in view of the fact
that the $5D$ Weyl multiplet decomposes into a $4D$ Weyl multiplet and an
additional vector multiplet. In spite of this, both in five and in
four dimensions, the matter multiplets are defined in a superconformal
background consisting only of the $5D$ or the $4D$ Weyl multiplet
fields, respectively. To fully establish this fact requires to also
consider the transformation rules beyond the linearized
approximation. As an aside we mention that a corresponding reduction
from four to three dimensions will involve a further extension of the
R-symmetry group. Namely, $\mathrm{SU}(2)\times\mathrm{U}(1)$ will
then be elevated to the group $\mathrm{SU}(2)\times\mathrm{SU}(2)$.

The central result of this paper will be to express the $5D$ off-shell
fields in terms of the $4D$ ones. We then verify that the $5D$
supersymmetric actions with terms quadratic in derivatives will yield
the $4D$ ones, at least for the bosonic fields. Subsequently we
consider the $5D$ action with terms quartic in derivatives
\cite{Hanaki:2006pj} and evaluate a number of characteristic terms in
the reduction to four dimensions to properly identify the $4D$
invariants that emerge. As it turns out the result decomposes into at
least three different invariants with higher derivatives. One of them
is the invariant based on a chiral superspace integral (the so-called
`F-term') \cite{Bergshoeff:1980is,LopesCardoso:2000qm}, another one
the invariant based on a full superspace integral (the so-called
`D-term') \cite{deWit:2010za}, and finally a (not necessarily
irreducible) invariant emerges that involves the square of the Ricci
tensor, which could in principle appear as an $N=2$
supercovariantization of the Gauss-Bonnet term.

These results enable us to address a number of issues that arose from
previous work on spinning BPS black holes in five dimensions and their
possible relation to four-dimensional black holes
\cite{Castro:2007hc,Castro:2007ci,Castro:2008ne,deWit:2009de}. In this
discussion the invariants with higher-derivative couplings play a
major role. Using a simplified model we find further support for the
results presented in \cite{deWit:2009de} and explain the subtle issues
for spinning black holes associated with the mixed gauge-gravitational
Chern-Simons term.

This paper is organized as follows. Section
\ref{sec:off-shell-dim-red-Weyl} presents the off-shell reduction to
four space-time dimensions of the five-dimensional Weyl multiplet, and
section \ref{sec:shell-dimens-reduct-matter} presents the
corresponding results for the vector multiplet and the
hypermultiplet. Section \ref{sec:five-four-dimens-fields-lagr} takes
into account the conversion of $5D$ symplectic Majorana spinors to the
$4D$ chiral spinor to obtain the explicit relations between $4D$ and
$5D$ fields. Subsequently these results are applied to supersymmetric
actions, leading to the precise decomposition of the $5D$ Lagrangians
into $4D$ supersymmetric Lagrangians. Section
\ref{sec:4D-5D-connection-bps-bh} addresses the situation regarding
BPS black holes, where differences were noted in the attractor
equations for the electric charges in the presence of
higher-derivative couplings. There are three appendices. Appendix
\ref{App:5-4D-Riemann-curv} discusses the relation between $5D$ and
$4D$ Riemann curvatures, the conversion of $5D$ spinors to $4D$ chiral
spinors is presented in appendix \ref{sec:spinors-four-five}, and some
$4D$ supersymmetry transformations are collected in appendix
\ref{App:SC}.

\section{Off-shell dimensional reduction; the Weyl multiplet}
\label{sec:off-shell-dim-red-Weyl}
\setcounter{equation}{0}
Starting from the superconformal transformations for $5D$
supermultiplets we will reduce the space-time dimension to $4D$ and
reinterpret the results in terms of the known superconformal
transformations in $4D$ dimensions. The first multiplet to consider
is the Weyl multiplet, because it acts as a background for other
supermultiplets, such as the vector multiplet and the
hypermultiplet. A second reason why the Weyl multiplet deserves
priority, is that it becomes reducible upon the reduction, unlike the
other (matter) supermultiplets. The Weyl multiplet in $D=5$ comprises
$32+32$ bosonic and fermionic degrees of freedom, which, in the
reduction to $D=4$ dimensions decomposes into the Weyl multiplet
comprising $24+24$ degrees of freedom, and a vector multiplet
comprising $8+8$ degrees of freedom. As we shall see, this
decomposition takes a subtle form off-shell. 

The independent fields of the Weyl multiplet of five-dimensional $N=1$
conformal supergravity consist of the f\"unfbein $e_M{}^A$, the
gravitino fields $\psi_M{}^i$, the dilatational gauge field $b_M$, the
R-symmetry gauge fields $V_{M i}{}^j$ (which is an anti-hermitean,
traceless matrix in the $\mathrm{SU}(2)$ indices $i,j$) and a tensor
field $T_{AB}$, a scalar field $D$ and a spinor field $\chi^i$. All
spinor fields are symplectic Majorana spinors. Our conventions are as
in \cite{deWit:2009de} apart from the supersymmetry parameters
$\epsilon^i$ which have been rescaled by a factor of two to bring the
normalization of the $5D$ supersymmetry algebra in line with the $4D$
algebra. The three gauge fields $\omega_M{}^{AB}$, $f_M{}^A$ and
$\phi_M{}^i$, associated with local Lorentz transformations, conformal
boosts and S-supersymmetry, respectively, are not independent as will
be discussed later. The infinitesimal Q, S and K transformations of
the independent fields, parametrized by spinors $\epsilon^i$ and
$\eta^i$ and a vector $\Lambda_\mathrm{K}{}^A$, respectively, are as
follows,\footnote{
  In five dimensions we consistently use world indices $M,N,\ldots$
  and tangent space indices $A,B,\ldots$. For fields that do not carry
  such indices the distinction between $5D$ and $4D$ fields may not
  always be manifest, but it will be specified in the text whenever
  necessary.} 
\begin{align}
  \label{eq:Weyl-susy-var}
  \delta e_M{}^A =&\,  \bar\epsilon_i \gamma^A \psi_M{}^i\,,
  \nonumber\\ 
  \delta \psi_{M}{}^i  =&\, 2\,
  {\cal  D}_M \epsilon^i + \tfrac1{2}\mathrm{i}\,
  T_{AB}( 3\,\gamma^{AB}\gamma_M-\gamma_M\gamma^{AB}) \epsilon^i
  -\mathrm{i}  \gamma_M\eta^i \,, \nonumber\\ 
  \delta V_{M i}{}^j =&\, 6 \mathrm{i}
  \bar\epsilon_{i} \phi_{M}{}^{j}
  -16 \bar\epsilon_{i}\gamma_M\chi^{j} -3 \mathrm{i}
  \bar\eta_{i}\psi_M{}^{j} + 
  \delta^i{}_j\,[-{3}\mathrm{i}\bar\epsilon_{k}\phi_{M}{}^{k}
  +8\bar\epsilon_{k}\gamma_M\chi^{k}+\ft{3}{2}\mathrm{i}
  \bar\eta_{k}\psi_M{}^{k}] \,, \nonumber \\
  \delta b_M =&\,
  \mathrm{i} \bar\epsilon_i\phi_M{}^i -4 \bar\epsilon_i\gamma_M
  \chi^i + \ft12\mathrm{i} \bar\eta_i\psi_M{}^i +2\Lambda _{K}{}^A
  e_{MA} \,, \nonumber\\
  \delta T_{AB} =&\,  \ft43 \mathrm{i} \bar\epsilon_i \gamma_{AB}
  \chi^i -\ft{1}{4} \mathrm{i} \bar\epsilon_i R_{AB}{}^i(Q)\,,
  \nonumber\\  
  \delta \chi^i =&\,  
  \ft 12 \epsilon^i D +\ft{1}{64} 
  R_{MN j}{}^{i}(V) \gamma^{MN} \epsilon^j 
  + \ft3{64}\mathrm{i}(3\, \gamma^{AB} \Slash{D}
  +\Slash{D}\gamma^{AB})T_{AB} \, \epsilon^i \nonumber\\
  &\,
  -\ft 3{16} T_{AB}T_{CD}\gamma^{ABCD}\epsilon^i 
  +\ft3{16} T_{AB}\gamma^{AB} \eta^i  \,, \nonumber\\
  \delta D =&\,
  2 \bar\epsilon_i \Slash{D} \chi^i - 2\mathrm{i}
  \bar\epsilon_i  T_{AB}\gamma^{AB} \chi^i - \mathrm{i}
  \bar\eta_i\chi^i \,. 
\end{align}
Under local scale transformations the various fields and
transformation parameters transform as indicated in table
\ref{tab:weyl-multiplet}. The derivatives $\mathcal{D}_M$ are
covariant with respect to all the bosonic gauge symmetries with the
exception of the conformal boosts. In particular we note
\begin{equation}
  \label{eq:D-epsilon}
\mathcal{D}_{M} \epsilon^i = \big( \partial_M - \tfrac{1}{4}
\omega_M{}^{CD} \, \gamma_{CD} + \tfrac1{2} \, b_M\big)
\epsilon^i + \tfrac1{2} \,{V}_{M j}{}^i \, \epsilon^j  \,, 
\end{equation}
where the gauge fields transform under their respective gauge
transformations according to
$\delta\omega_M{}^{AB}=\mathcal{D}_M\lambda^{AB}$, $\delta b_M=
\mathcal{D}_M\Lambda_D$ and $\delta V_{M i}{}^j= \mathcal{D}_M
\Lambda_i{}^j$, with $(\Lambda_i{}^j)^\ast\equiv \Lambda^i{}_j=
- \Lambda_j{}^i$. The derivatives $D_M$ are covariant with
respect to all the superconformal symmetries. 
  
%
\begin{table}[t]
\centering
\begin{tabular}{|c|ccccccc|ccc|ccc|} 
\hline 
 & &\multicolumn{8}{c}{Weyl multiplet} & &
 \multicolumn{2}{c}{parameters} & \\  \hline 
 field & $e_M{}^{A}$ & $\psi_M{}^i$ & $b_M$ &
 ${V}_{M\,i}{}^j$ & $T_{AB} $ & 
 $ \chi^i $ & $D$ & $\omega_{M}{}^{AB}$ & $f_M{}^A$ &$\phi_M{}^i$&
 $\epsilon^i$ & $\eta^i$  
 & \\ \hline
$w$  & $-1$ & $-\tfrac12 $ & 0 &  0 & 1 & $\tfrac{3}{2}$ & 2 & 0 &
1 & $\tfrac12 $ & $-\tfrac12$  & $\tfrac12$ & \\ \hline 
\end{tabular}
\vskip 2mm
\renewcommand{\baselinestretch}{1}
\parbox[c]{14.8cm}{\caption{\footnotesize
    Weyl weights $w$ of the 
    Weyl multiplet component fields and the supersymmetry
    transformation parameters. \label{tab:weyl-multiplet}}}    
\end{table}

The above supersymmetry variations and also the conventional
constraints that we have to deal with in due time, depend on a number
of supercovariant curvature tensors, which take the following form,
\begin{align}
  \label{eq:curvatures}
  R(P)_{MN}{}^A =&\, 2\,\mathcal{D}_{[M}e_{N]}{}^A
  -\tfrac12 \bar\psi_{M i}\gamma^A \psi_{N}{}^i
  \,,\nonumber\\[.4ex]   
  R(M)_{MN}{}^{AB} =&\, 2\,\partial_{[M}\omega_{N]}{}^{AB}
  -2\, \omega_{[M}{}^{AC} \omega_{N]C}{}^{B} -8\, e_{[M}{}^{[A}
  f_{N]}{}^{B]} + \mathrm{i} \bar\psi_{[M
  i}\gamma^{AB}\phi_{N]}{}^i \nonumber\\
  &\, - \tfrac14 \mathrm{i} T^{CD}
  \,\bar\psi_{[M i}\big(6\gamma^{[A}\gamma_{CD}\gamma^{B]}
  -\gamma^{AB}\gamma_{CD}-\gamma_{CD}\gamma^{AB}\big)\psi_{N]}{}^i
  \nonumber\\ 
  &\,
  -\tfrac12\bar\psi_{[M i}(\gamma_{N]} R^{ABi}(Q) + 2\,\gamma^{[A}
  R_{N]}{}^{B]i}(Q)) + 8\,
  e_{[M}{}^{[A}\,\bar\psi_{N]i}\gamma^{B]}\chi^i \,,\nonumber\\[.4ex]  
  R(D)_{MN} =&\, 2\,\partial_{[M} b_{N]}-4\,f_{[M}{}^A
  e_{N]A}   
  - \mathrm{i} \bar{\psi}_{[M i} \phi_{N]}{}^i + 4\,\bar\psi_{[M
  i}\gamma_{N]}\chi^i \,. \nonumber\\[.4ex]  
  R(V)_{MN i}{}^j =&\, 2\, \mathcal{\partial}_{[M}V_{N]i}{}^j
  - V_{[M i}{}^k V_{N] k}{}^j   \nonumber \\
  &\,
  - 6 \mathrm{i}\,\bar\psi_{[M i} \phi_{N]}{}^j +16 
  \bar\psi_{[M i}\gamma_{N]} \chi^j
  +\delta_i{}^j\big[3 \mathrm{i}\,\bar\psi_{[M k} \phi_{N]}{}^k -8 
  \bar\psi_{[M k}\gamma_{N]} \chi^k \big] \,,\nonumber\\[.4ex]  
  R(Q)_{MN}{}^i =&\, 2\, \mathcal{D}_{[M}\psi_{N]}{}^i -2
  \mathrm{i}\, 
  \gamma_{[M} \phi_{N]}{}^i  +\ft12\mathrm{i}\, T_{AB}\big(
  3\,\gamma^{AB}\gamma_{[M} -\gamma_{[M}\gamma^{AB}\big)
  \psi_{N]}{}^i  \,. 
\end{align}
The curvature tensor $R_{MN}{}^A(K)$ associated with the conformal
boosts has not been defined and is not needed henceforth. The
curvature tensor $R_{MN}{}^i(S)$ will be discussed shortly.

The conventional constraints are as follows, 
\begin{align}
  \label{eq:conv-constraints-5}
  R(P)_{MN}{}^A =&\, 0\,,\nonumber \\
  \gamma^M R(Q)_{MN}{}^i =&\, 0\,,\nonumber\\
  e_A{}^M\, R(M)_{MN}{}^{AB} =&\, 0 \,. 
\end{align}
These conditions determine the gauge fields $\omega_M{}^{AB}$,
$f_M{}^A$ and $\phi_M{}^i$. The conventional constraints lead to
additional constraints on the curvatures when combined with the
Bianchi identities. In this way one derives $R(M)_{[ABC]D} =0=
R(D)_{AB}$ and the pair-exchange property $R(M)_{ABCD}=R(M)_{CDAB}$
from the first and the third constraint.  The second constraint, which
implies also that $\gamma_{[MN} R(Q)_{PQ]}{}^i =0$, determines the
curvature $R(S)_{MN}{}^i$, which we refrained from defining
previously. It turns out to be proportional to $R(Q)_{MN}{}^i$ and
derivatives thereof,
\begin{align}
  \label{eq:R(S)} 
  R(S)_{MN}{}^i =&\, -\mathrm{i} \Slash{D} 
  R(Q)_{MN}{}^i - {\rm i} \gamma_{[M} 
  D^{P}  R(Q)_{N]P}{}^i
  - 4\,\gamma_{MN} T^{PQ} R(Q)_{PQ}{}^i  \nonumber\\
  &\,  +18\, T^{PQ} \gamma_Q \gamma_{[M} 
  R(Q)_{N]P}{}^i -5 \, T^{PQ}\gamma_{PQ}  
  R(Q)_{MN}{}^i  - 12\, T^P{}_{[M} R(Q)_{N]P}{}^i \,.
\end{align}

The reduction to four space-time dimensions is effected by first
carrying out the standard Kaluza-Klein decompositions on the various
fields, to guarantee that the resulting $4D$ fields will transform
consistently under four-dimensional diffeomorphisms. The space-time
coordinates are decomposed into $x^M\to (x^\mu,x^5)$, where $x^5$
denotes the fifth coordinate that will be suppressed in the
reduction. Subsequently the vielbein field and the dilatational
gauge field are then written in special form, by means of an
appropriate local Lorentz transformation and a conformal boost,
respectively. In obvious notation,
\begin{equation}
  \label{eq:kk-ansatz}
  e_M{}^A= \begin{pmatrix} e_\mu{}^a &  B_\mu\phi^{-1} \\[4mm]
    0 & \phi^{-1}
    \end{pmatrix} \;,\qquad
    e_A{}^M= \begin{pmatrix} e_a{}^\mu & - e_a{}^\nu B_\nu \\[4mm]
    0 & \phi
    \end{pmatrix}\;,\qquad
    b_M = \begin{pmatrix} b_\mu \\[4mm]  0
  \end{pmatrix} \,.
\end{equation}
On the right-hand side of these decompositions, we exclusively used
four-dimensional notation, with world and tangent-space indices, 
$\mu,\nu,\ldots$ and $a,b,\ldots$, taking four values. Observe that
the scaling weights for $e_M{}^A$ and $e_\mu{}^a$ are equal to $w=-1$,
while for $\phi$ we have $w=1$. The fields $b_M$ and $b_\mu$ have
weight $w=0$. The above formulae suffice to express the $5D$ Riemann
curvature tensor in terms of the $4D$ Riemann tensor and the fields
$\phi$ and $B_\mu$. The corresponding equations are collected in
appendix \ref{App:5-4D-Riemann-curv} and will be needed later on. 

We now turn to the supersymmetry transformations. Since we have
imposed gauge choices on the vielbein field and the dilatational gauge
field, one has to include compensating Lorentz and special conformal
transformations when deriving the $4D$ Q-supersymmetry transformations
to ensure that the gauge conditions are preserved. Only the parameter
of the Lorentz transformation is relevant, and it is given by,
\begin{equation}
  \label{eq:comp-Lor}
  \varepsilon^{a5} = -\varepsilon^{5a} = - \phi\,
  \bar\epsilon_i\gamma^a\psi^i \,,
\end{equation}
where we assumed the standard Kaluza-Klein decomposition on the
gravitino fields, 
\begin{equation}
  \label{eq:gravitino-KK}
  \psi_M{}^i =  \begin{pmatrix}\psi_\mu{}^i+ B_\mu \psi^i\\[4mm]
    \psi^i \end{pmatrix}\;, 
\end{equation}
which ensures that $\psi_\mu{}^i$ on the right-hand side transforms as
a $4D$ vector.  Upon including this extra term, one can write down the
Q- and S-supersymmetry transformations on the $4D$ fields defined
above. As a result of this, the $4D$ and $5D$ supersymmetry
transformation will be different. For instance, the supersymmetry
transformations of the $4D$ fields $e_\mu{}^a$, $\phi$ and $B_\mu$
read,
\begin{align}
  \label{eq:susy-e-B-phi}
  \delta e_\mu{}^a =&\,  \bar\epsilon_i\gamma^a\psi_\mu{}^i
  \,, \nonumber\\[.2ex]
  \delta\phi =&\, - \phi^2\,\bar\epsilon_i\gamma^5\psi^i\,,
  \nonumber \\[.2ex]
  \delta B_\mu=&\,  \phi^2 \,\bar\epsilon_i\gamma_\mu\psi^i +
  \phi \,\bar\epsilon_i\gamma^5\psi_\mu{}^i \,,
\end{align}
where the first term in $\delta B_\mu$ originates from the
compensating transformation \eqref{eq:comp-Lor}. Consequently the
supercovariant field strength of $B_\mu$ contains a term that is not
contained in the supercovariant five-dimensional curvature
$R(P)_{MN}{}^A$. Therefore the $5D$ spin-connection components are not
supercovariant with respect to $4D$ supersymmetry, as is shown below,
\begin{align}
  \label{eq:spin-connection}
  \omega_M{}^{ab} =&\, \begin{pmatrix} \omega_\mu{}^{ab} \\[4mm] 
    0 \end{pmatrix} + \tfrac12 \phi^{-2} \hat F(B)^{ab} \, 
  \begin{pmatrix} B_\mu \\[4mm] 1\end{pmatrix} \;, \nonumber \\[.8ex]
  \omega_M{}^{a5} =&\, -\tfrac12 \begin{pmatrix} \phi^{-1}
    \hat F(B)_\mu{}^a +\phi\,\bar\psi_{\mu i}\gamma^a\psi^i \\[4mm]
    0 \end{pmatrix} -\phi^{-2} D^{a}\phi  \,    
  \begin{pmatrix} B_\mu \\[4mm] 1\end{pmatrix} \;.
\end{align}
Here we introduced the supercovariant field strength and derivative
(with respect to $4D$ supersymmetry), 
\begin{align}
  \label{eq:supercov-FB-Dphi}
  \hat F(B)_{\mu\nu} =&\, 2\,\partial_{[\mu} B_{\nu]} - \phi^2\,
  \bar\psi_{[\mu i}\gamma_{\nu]} \psi^i -\tfrac12 \phi\,\bar\psi_{\mu
    i}  \psi_{\nu}{}^i \,,\nonumber\\
  D_\mu\phi =&\, (\partial_\mu -b_\mu) \phi + \tfrac12 \phi^2
  \,\bar\psi_{\mu i}\gamma_5 \psi^i \,.  
\end{align}

Subsequently we write down corresponding Kaluza-Klein decompositions
for some of the other fields of the Weyl multiplet, which do not
require special gauge choices,
\begin{equation}
  \label{eq:1Weyl-KK}
    {V}_{M i}{}^j=
  \begin{pmatrix}{V}_{\mu i}{}^j+ B_\mu
    {V}_{i}{}^j\\[4mm] 
    {V}_{i}{}^j \end{pmatrix}\;,\qquad
  \phi_M{}^i =  \begin{pmatrix}\phi_\mu{}^i+ B_\mu \phi^i\\[4mm]
    \phi^i \end{pmatrix}\;\qquad 
   T_{AB} = \begin{pmatrix} T_{ab} \\[4mm]  T_{a5}\equiv -\tfrac16
    A_a \end{pmatrix}  \,. 
\end{equation}
For the fermions there is yet no need to introduce new notation,
because the spinors have an equal number of components in five and
four space-time dimensions.  Eventually we will convert to standard
four-dimensional chiral spinors.

Hence we are now ready to consider the supersymmetry transformations
of the spinor fields originating from the $5D$ gravitino fields. Up to
possible higher-order spinor terms, one derives from
\eqref{eq:Weyl-susy-var},
\begin{align}
  \label{eq:susy-W-gravitino}
  \delta\psi^i =&\, -\tfrac12
  \phi^{-2} \big[\hat F(B)_{ab}- \mathrm{i}\gamma_5 \phi (
  3\,T_{ab} -\tfrac14 \mathrm{i} \phi^{-1} \hat F(B)_{ab}\gamma_5
  )\big]   \gamma^{ab}\epsilon^i \nonumber\\
  &\,
  +\phi^{-2}\big[ \Slash{D} \phi \,\gamma^5 - \mathrm{i} 
  \Slash{A}\phi\big] \epsilon^i -
  {V}^i{}_j \epsilon^j \nonumber\\[.2ex]
  &\,
    - \mathrm{i}  \gamma_5 \phi^{-1} \big[ \eta^i
    +\tfrac13\Slash{A}\gamma_5\epsilon^i  +\tfrac1{8} \mathrm{i} 
  \gamma_5\phi^{-1}(F(B)_{ab}-4\mathrm{i}\phi
   T_{ab}\gamma_5)\gamma^{ab} \epsilon^i\big]  \,, \nonumber\\[.2ex] 
  \delta\psi_\mu{}^i =&\,2\,\big(\partial_\mu
  -\tfrac14\omega_\mu{}^{ab}\gamma_{ab}+\tfrac12 b_\mu
  +\tfrac12\mathrm{i} e_\mu{}^a \,A_a \gamma_5 \big)\epsilon^i
  +  {V}_{\mu j}{}^i \epsilon^j \nonumber\\
  &\, + \tfrac12 \mathrm{i}\big[3\, T_{ab} - \tfrac14 \mathrm{i}
  \phi^{-1} \hat F(B)_{ab}\gamma_5  \big] \gamma_{ab} \gamma_\mu
  \epsilon^i  \nonumber\\
  &\,
  - \mathrm{i} \gamma_\mu \big[\eta^i
   +\tfrac13\Slash{A}\gamma_5 \epsilon^i  +\tfrac1{8}
   \mathrm{i} \gamma_5\phi^{-1}(\hat F(B)_{ab}-4\mathrm{i}\phi
   T_{ab}\gamma_5)\gamma^{ab} \epsilon^i \big]\,.  
   \end{align}
Although this result is not yet complete, it already exhibits some of
the systematic features that will turn out to be universal. Therefore
let us have a brief perusal of these initial results. 

The fields whose transformations we have determined will belong to two
$4D$ supermultiplets, namely the Weyl and the Kaluza-Klein vector
multiplet. Clearly, the fields $e_\mu{}^a$ and $\psi_\mu{}^i$ belong
to the Weyl multiplet, whereas $\phi$, $B_\mu$ and $\psi^i$ belong to
the vector multiplet. Their transformations shown in
\eqref{eq:susy-e-B-phi} and \eqref{eq:susy-W-gravitino} have many
features in common with the standard $4D$ transformations of a Weyl
and a vector multiplet. An obvious puzzle is the fact that we have
identified only one real scalar, whereas the $D=4$ vector multiplet
contains a complex scalar. Furthermore, we note that the field $A_a$
seems to play the role of a $\mathrm{U}(1)$ gauge field, because it
appears to covariantize the derivatives on $\phi$ and $\epsilon^i$ in
\eqref{eq:susy-W-gravitino}, in spite of the fact that it is actually
an auxiliary field in $D=5$. As we shall see in due course, the
resolution of these two problems is related.

Another observation is that a particular linear combination of the
$5D$ tensor components $T_{ab}$ and the supercovariant field strength
$\hat F(B)_{ab}$ appears in the transformations
\eqref{eq:susy-W-gravitino} in precisely the same form as the $4D$
auxiliary tensor $T_{ab}$, suggesting that the latter is not just
proportional to the corresponding $5D$ tensor field components. The
same combination will also appear in other transformation rules, as we
shall see in, for instance, section
\ref{sec:shell-dimens-reduct-matter}. Finally, S-supersymmetry
transformations are accompanied by another universal combination of
$T_{ab}$ and $\hat F(B)_{ab}$. Obviously such a field-dependent
component in the S-supersymmetry transformation can be dropped
provided that it appears universally for all other fields, as it can
be absorbed into $\eta^i$.

As it turns out this pattern becomes more complicated when including
terms of higher order in the fermions. Apart from new contributions to
the expressions noted above, it turns out that also R-symmetry will
appear on the right-hand side with parameters that involve the spinors
$\psi^i$. Again this R-symmetry transformation acts universally on all
the fields. Hence the conclusion is that the $5D$ supersymmetries
decompose under the reduction into the $4D$ supersymmetries up to $4D$
field-dependents S-supersymmetries and $\mathrm{SU}(2)$
R-symmetries. This property explains why only a careful analysis can
reveal how the off-shell supermultiplets reduce to lower dimension, as
precise knowledge of this decomposition is required before one can
reliably extract the $4D$ transformations. These transformations will
then subsequently identify the $4D$ fields in terms of the $5D$ ones
(up to straightforward calibrations). After verifying that the
decomposition is universally realized these extra symmetries with
field-dependent coefficients can be dropped.

However, there is yet another surprise, as we will discover the
presence of a chiral $\mathrm{U}(1)$ transformation in the
supersymmetry variations with a universal coefficient. Since chiral
$\mathrm{U}(1)$ does not constitute a symmetry of the $5D$ theory, the
contribution from this transformation cannot be dropped and should be
kept until the end. We will discuss its fate in due time. Obviously
these transformations will play a crucial role in extending the
R-symmetry to $\mathrm{SU}(2)\times\mathrm{U}(1)$.

Summarizing, we intend to first establish that the dimensional
reduction of the $5D$ supersymmetry variations, according to the
procedure sketched above, takes the form of the $4D$ supersymmetry 
variations combined with a field-dependent S-supersymmetry
transformation, a field-dependent $\mathrm{SU}(2)$ R-symmetry
transformation, and a field-dependent $\mathrm{U}(1)$ chiral
transformation,
\begin{equation}
  \label{eq:D5-D4-decomp}
  \delta_\mathrm{Q}(\epsilon)\big|^\mathrm{reduced}_{5D} \Phi =
  \delta_\mathrm{Q}(\epsilon)\big|_{4D} \Phi + 
  \delta_\mathrm{S}(\tilde\eta)\big|_{4D} \Phi +
  \delta_{\mathrm{SU}(2)}(\tilde\Lambda)\big|_{4D}\Phi  +
  \delta_{\mathrm{U}(1)}(\tilde\Lambda^0)\big|_{4D}\Phi\,.  
\end{equation}
To give a meaning to the right-hand side one has to identify fields
$\Phi$ that transform covariantly in the $4D$ setting, so that all
transformations in the above decomposition are clearly defined. The
identification of these fields is done iteratively. Here
one has to realize that the $5D$ transformations for the Weyl
multiplet are defined in a background consisting of the $5D$ Weyl
multiplet, whereas the $4D$ transformations of the matter multiplets
are defined in the $4D$ background. But the field-dependent parameters
in \eqref{eq:D5-D4-decomp} are not restricted and still depend on a
variety of the $5D$ Weyl multiplet fields.  These parameters,
$\tilde\eta^i$, $\tilde \Lambda$ and $\tilde \Lambda^0$, are defined
as follows (consistent with the lowest-order contributions that we
have already exhibited in \eqref{eq:susy-W-gravitino}),
\begin{align}
  \label{eq:field-dep-S-R}
  \tilde\eta^i=&\, \tfrac13 \Slash{A}\,\gamma^5 \epsilon^i  +\tfrac1{8}
   \mathrm{i} \gamma^5\phi^{-1}(\hat F(B)_{ab}-4\mathrm{i}\phi\,
   T_{ab}\gamma_5)\gamma^{ab} \epsilon^i   \nonumber \\
   &\, +\tfrac1{4}\mathrm{i} \phi^{2}
  \big(\bar\psi_j\gamma^5\psi^i\,\gamma_5 -\bar\psi_j \psi^i
    +\bar\psi_j\gamma^a\psi^i\, \gamma_a 
  +\tfrac{1}{2} \bar\psi_k\gamma^5\gamma^a\psi^k\, \gamma_5\gamma_a \,
  \delta_j{}^i \big) \epsilon^j\,,  \nonumber\\[.2ex] 
  \tilde\Lambda_i{}^j  =&\, \phi\,\big( \bar\epsilon_k
  \gamma^5 \psi^l \,\varepsilon_{il}\varepsilon^{jk} - \tfrac12
  \bar\epsilon_k \gamma^5 \psi^k\,\delta_i{}^j  \big) \,, \nonumber 
  \\[.2ex]
  \tilde\Lambda^0=&\,\mathrm{i} \phi \, \bar \epsilon_i
  \psi^i\,.  
\end{align}

Let us briefly discuss the non-linear corrections to
\eqref{eq:susy-W-gravitino}, whose contributions were already included
in \eqref{eq:field-dep-S-R}. They originate from three sources, namely
the compensating Lorentz transformation \eqref{eq:comp-Lor}, the
non-supercovariant term in the spin connection $\omega_\mu{}^{a5}$,
and the non-linearity in the definition of the $4D$ gravitini
$\psi_\mu{}^i$ in terms of the $5D$ fields
(c.f. \eqref{eq:gravitino-KK}). Concentrating on variations that
explicitly contain $\psi_\mu{}^i$, one easily notes that they no
longer satisfy the standard supercovariance properties (which are
manifest in four and five dimensions). In principle it is possible to
absorb some of the unwanted terms in some of the bosonic fields
appearing on the right-hand side of \eqref{eq:susy-W-gravitino} or in
the field-dependent S-transformations. The only fields, however, that
can accommodate terms proportional to the bare (i.e. not contained in
covariant objects) gravitini, are the R-symmetry gauge fields $V_{\mu
  i}{}^j$. However, in that case the supersymmetry variation of these
gauge fields will acquire terms proportional to derivatives on the
supersymmetry parameter $\epsilon^i$, which can only be interpreted as
an extra field-dependent $\mathrm{SU}(2)$ R-symmetry transformations,
as is already indicated in \eqref{eq:D5-D4-decomp}. However, there are
also higher-order variations proportional to $\psi^i$, so the
situation becomes considerably more involved. In deciding how to deal
with all these terms, some guidance can be obtained from reducing, at
the same time, the matter multiplets. But for the sake of clarity, we
prefer not to mix the presentation of the Weyl multiplet reduction
with the presentation of the reduction of the matter multiplets. The
latter will therefore be postponed to section
\ref{sec:shell-dimens-reduct-matter}. At this point we will simply
take note this extra evidence and restrict our discussion here to the
Weyl multiplet reduction.

The result of the reduction motivates the following redefinitions of
the various fields, 
\begin{align}
  \label{eq:field-redef}
  \hat A_\mu =&\, A_a \,e_\mu{}^a -\tfrac12\mathrm{i}\, \phi\,
  \bar\psi_j\psi_\mu{}^j-\tfrac1{4}\mathrm{i}\, \phi^2\,
  \bar\psi_j\gamma^5\gamma_\mu\psi^j \,, \nonumber \\[.2ex]
  \hat{T}_{ab}=&\, 24\,T_{ab} + \mathrm{i} \phi^{-1}\,\varepsilon_{abcd}\,
  \hat F(B)^{cd} 
   -\mathrm{i} \phi^{2} \bar\psi_i\gamma_{ab} \psi^i
   \,,\nonumber \\[.2ex] 
   \hat V_j{}^i=&\, \phi^2\big(V_j{}^i - \tfrac32 \phi\,
   \bar\psi_j\,\gamma^5\psi^i\big) \,,\nonumber\\ 
   \hat V_\mu{}_j{}^i =&\, {V}_\mu{}_j{}^i
  - \phi \big( \bar\psi_{\mu j} \gamma^5 \psi^i
   - \tfrac12 \delta_j{}^i\,
  \bar\psi_{\mu k} \gamma^5 \psi^k  \big)
  -\tfrac12  \phi^2\, \bar\psi_{j} \gamma_\mu \psi^i\,.
\end{align}
These are the linear combinations that emerge in the $4D$
supersymmetry transformations. Their S-supersymmetry transformations
turn out to be relevant and we note the following result,
\begin{align}
  \label{eq:S-ATV}
   \delta\hat A_\mu=&\,\tfrac12\bar\psi_{\mu j}\gamma^5 \eta^j\,,
   \nonumber \\[.2ex] 
   \delta\hat T_{ab} =&\, 0\,, \nonumber \\[.2ex]
   \delta \hat V_j{}^i=&\, 0\,,  \nonumber\\[.2ex]
   \delta\hat V_{\mu j}{}^i=&\,  -2\mathrm{i}\big(\bar\psi_{\mu j}\,\eta^i
   -\tfrac1{2}\delta_j{}^i\,\bar\psi_{\mu k}\,\eta^k \big) \,. 
\end{align}
In particular, note that the factor in the variation of $\hat V_{\mu
  i}{}^j$ has now changed as compared to the factor that appears in
the corresponding $5D$ S-variation given in
\eqref{eq:Weyl-susy-var}. Furthermore, note that $\hat A_\mu$ is not
supercovariant; its Q-supersymmetry variation contains a term
proportional to the derivative of the supersymmetry
parameter. Eventually $\hat A_\mu$ will be related to a gauge field
associated with the $4D$ $\mathrm{U}(1)$ R-symmetry. This is consistent
with the fact that $\hat A_\mu$ transforms into the gravitino fields
under S-supersymmetry, in agreement with $4D$ results.

With these notational changes we repeat the Q-supersymmetry
transformations of \eqref{eq:susy-e-B-phi} and
\eqref{eq:susy-W-gravitino}, including also the higher-order
contributions. First we consider those associated with the
Kaluza-Klein vector multiplet,
\begin{align}
  \label{eq:susy-KK}
    \delta\phi =&\, -\phi^2\,\bar\epsilon_i\gamma^5\psi^i\,,
  \nonumber \\[.2ex]
  \delta B_\mu=&\, \phi^2 \,\bar\epsilon_i\gamma_\mu\psi^i +
  \phi \,\bar\epsilon_i\gamma^5\psi_\mu{}^i \,,\nonumber
  \\[.2ex]  
  \delta\big(\phi^2\psi^i\big) =&\, -\tfrac12 \big[\hat
  F(B)_{ab}-\tfrac18 \mathrm{i}\gamma_5 \phi\, \hat T_{ab} \big]
  \gamma^{ab}\epsilon^i  \nonumber\\
  &\, - \big[(\partial_\mu-b_\mu)\phi\,\gamma^5
  +\tfrac12\phi^2(\bar\psi_{\mu j} 
  \gamma^5\psi^j\,\gamma^5 +\bar\psi_{\mu j}\psi^j) +\mathrm{i}\phi\, \hat
    A_\mu \big]\gamma^\mu \epsilon^i \nonumber\\
    &\, + {\hat V}_j{}^i \epsilon^j
  -\tfrac12 \mathrm{i}\tilde\Lambda^0 \phi^2
  \gamma^5\psi^i  -\mathrm{i}\phi \gamma^5 \eta^i\,, 
\end{align}
where here and henceforth we suppress the S-supersymmetry and
R-symmetry transformations proportional to the field-dependent
parameters $\tilde\eta^i$ and $\tilde\Lambda_i{}^j$. However, we did
include the $\mathrm{U}(1)$ transformation with parameter
$\tilde\Lambda^0$, just as in the next formula. Apart from some minor
details, these variations show the same structure as the $4D$
transformation rules of vector multiplet, except that we have been
unable to identify the second scalar field. The field $\hat V_j{}^i$
is obviously related to the auxiliary field of this vector multiplet.

The Q-supersymmetry transformations of the Weyl multiplet fields are
as follows, 
\begin{align}
  \label{eq:susy-weyl1}
  \delta e_\mu{}^a =&\,  \bar\epsilon_i\gamma^a\psi_\mu{}^i
  \,, \nonumber\\[.2ex]
    \delta\psi_\mu{}^i =&\,2\,\big(\partial_\mu
  -\tfrac14\omega_\mu{}^{ab}\gamma_{ab}+\tfrac12 b_\mu
  +\tfrac12\mathrm{i}  \hat A_\mu \gamma_5 \big)\epsilon^i
  +  {\hat V}_{\mu j}{}^i \epsilon^j \nonumber\\
  &\, + \tfrac1{16} \mathrm{i} \hat T_{ab} \gamma^{ab} \gamma_\mu
  \epsilon^i  - \tfrac12\mathrm{i}\tilde\Lambda^0 \,\gamma^5 \psi_\mu{}^i
  -\mathrm{i} \gamma_\mu \eta^i \,, 
   \end{align}
Also in this case the variations show a close resemblance to the $4D$
transformation rules of the $4D$ Weyl multiplet fields, with $\hat
V_{\mu j}{}^i$ playing the role of the $\mathrm{SU}(2)$ gauge
fields. In both the above results $\hat A_\mu$ seems to play the role
of the $\mathrm{U}(1)$ chiral gauge field, and $\hat T_{ab}$ is the
$4D$ tensor field. 

We now return to the issue of the missing spinless field in the
Kaluza-Klein vector multiplet. The crucial observation is that the
expressions obtained so far are consistent with the assumption that we
are dealing with a gauge-fixed version of the theory. So we simply
have to introduce a phase for the vector multiplet scalar which
transforms locally under $\mathrm{U}(1)$ transformations. This is
achieved by introducing the following R-covariant spinors,
transforming under local $\mathrm{SU}(2)\times\mathrm{U}(1)$
R-symmetry transformations,
\begin{align}
  \label{eq:compensating-chiral-tr}
  \epsilon^i\vert^\mathrm{Rcov} =&\,
  \exp[-\tfrac12\mathrm{i}\varphi\,\gamma^5]\, 
  \epsilon^i\,,\nonumber \\ 
  \eta^i\vert^\mathrm{Rcov} =&\,
  \exp[\tfrac12\mathrm{i}\varphi\,\gamma^5]\,\eta^i \,,\nonumber\\ 
  \psi_\mu{}^i\vert^\mathrm{Rcov} =&\,
  \exp[-\tfrac12\mathrm{i}\varphi\,\gamma^5]\,\psi_\mu{}^i\,,\nonumber\\ 
  \psi^i\vert^\mathrm{Rcov} =&\,
  \exp[-\tfrac12\mathrm{i}\varphi\,\gamma^5]\, \psi^i 
\end{align}
and assume that the phase factor $\varphi$ transforms under
supersymmetry and under a new local $\mathrm{U}(1)$ group according
to
\begin{equation}
  \label{eq:transf-varphi}
  \delta\varphi= \Lambda^0 - \mathrm{i} \phi\,\bar\epsilon_i
  \psi^i \,,
\end{equation}
where $\Lambda^0$ is now an arbitrary space-time dependent function.
Imposing a $\mathrm{U}(1)$ gauge choice $\varphi= 0$ then generates a
compensating $\mathrm{U}(1)$ component in the Q-variations, so that
these terms re-emerge in the supersymmetry transformations for the
fermions. The R-covariant spinors are not yet converted to the
standard chiral $4D$ spinors, but possess already all the necessary
features 

The variations \eqref{eq:susy-KK} can now be rewritten in terms of
R-covariant spinors. Here and henceforth we will suppress the
superscript $\scriptstyle\mathrm{Rcov}$. Furthermore we employ chiral
spinors defined by $\gamma^5\psi_\pm=\pm \psi_\pm$. The result takes
the form,
\begin{align}
  \label{eq:susy-KK-u1}
    \delta\big(\mathrm{e}^{\mp\mathrm{i}\varphi}\, \phi\big)  =&\, \mp2
    \phi^2\,\bar\epsilon_i \psi^i_\pm\,, 
  \nonumber \\[.2ex]
  \delta B_\mu=&\, 
  \bar\epsilon_i\big[\gamma_\mu\,\phi^2 \psi^i_- +
   \phi\,\mathrm{e}^{\mathrm{i}\varphi} \psi_{\mu}{}^i_+\big]  +
    \bar\epsilon_i\big[\gamma_\mu\,\phi^2 \psi^i_+
  - \phi\,\mathrm{e}^{-\mathrm{i}\varphi}
  \,\psi_{\mu}{}^i_-\big]\,,\nonumber 
  \\[.2ex]  
  \delta\big(\phi^2\psi^i_\pm\big) =&\, -\tfrac12 \big[\hat
  F(B)_{ab}\mp \tfrac18 \mathrm{i} \phi\, \hat T_{ab} \big]
  \gamma^{ab}\epsilon^i_\pm  \nonumber\\
  &\, \mp  \Slash{\hat D} 
    (\phi\,\mathrm{e}^{\mp\mathrm{i}\varphi} )\, 
  \epsilon^i_\mp + {\hat V}_j{}^i \epsilon^j_\pm 
    \mp\mathrm{i}\phi\,\mathrm{e}^{\mp\mathrm{i}\varphi}\eta^i_\pm\,, 
\end{align}
where $\hat D_\mu(\phi\,\mathrm{e}^{\mp\mathrm{i}\varphi})$ is a
supercovariant and $\mathrm{U}(1)$ covariant derivative, defined by
\begin{equation}
  \label{eq:superco-u1-der} 
  \hat D_\mu(\phi\,\mathrm{e}^{\mp\mathrm{i}\varphi})= 
  (\partial_\mu-b_\mu) (\phi\,\mathrm{e}^{\mp\mathrm{i}\varphi})  \pm\phi^2
  \bar\psi_{\mu i}\psi^i_\pm \pm\mathrm{i}  A_\mu
  (\phi\,\mathrm{e}^{\mp\mathrm{i}\varphi}) \,. 
\end{equation}
Here the $\mathrm{U}(1)$ connection equals
\begin{equation}
  \label{eq:def-A}
 A_\mu= \hat A_\mu + \partial_\mu\varphi\,.
\end{equation}
Hence the R-symmetry group has now been extended to
$\mathrm{SU}(2)\times\mathrm{U}(1)$.  Observe that $A_\mu$ transforms
covariantly under supersymmetry. The definition \eqref{eq:def-A} can
be written in supercovariant form,
\begin{equation}
  \label{eq:sc-def-A}
  \hat D_\mu\varphi = -6\,T_{a5} \,e_\mu{}^a +\tfrac14 \phi^2
  \bar\psi_i\gamma^5\gamma_\mu\psi^i \,. 
\end{equation}

The same manipulations can be applied to the fields of the Weyl
multiplet, and we give the result for the vielbein and gravitino
fields, the latter again written with R-covariant spinors,
\begin{align}
  \label{eq:susy-weyl2}
  \delta e_\mu{}^a =&\,  \bar\epsilon_i\gamma^a\psi_\mu{}^i_+
  + \bar\epsilon_i\gamma^a\psi_\mu{}^i_-
  \,, \nonumber\\[.2ex]
    \delta\psi_\mu{}^i_\pm =&\,2\,\big(\partial_\mu
  -\tfrac14\omega_\mu{}^{ab}\gamma_{ab}+\tfrac12 b_\mu
  \pm\tfrac12\mathrm{i} A_\mu  \big)\epsilon^i_\pm
  + {\hat V}_{\mu j}{}^i \epsilon^j_\pm \nonumber\\
  &\, + \tfrac1{16} \mathrm{i}  \mathrm{e}^{\mp\mathrm{i}\varphi} \hat
  T_{ab} \gamma^{ab} \gamma_\mu 
  \epsilon^i_\mp  -\mathrm{i} \gamma_\mu \eta^i_\mp \,, 
\end{align}

Apart from different spinor conventions and normalizations the
supersymmetry variations take the same form as the corresponding ones
in four dimensions. There is a subtlety, which is that the $4D$ tensor
field can be split in selfdual and anti-selfdual components and should
be identified with $\mathrm{e}^{\pm\mathrm{i} \varphi} \, \hat
T_{ab}{}^\pm$, where the superscript $\pm$ on the tensor indicates its
duality phase. After this identification the correspondence between
the $\mathrm{U}(1)$ and the chirality/duality assignments is precisely
as in the four-dimensional theory. The Kaluza-Klein multiplet
transforms as a proper vector supermultiplet in a $4D$ superconformal
background. Its auxiliary field $\hat V_i{}^j$ is indeed an
$\mathrm{SU}(2)$ vector. As a check we have verified that its
transformations take the form expected from the $4D$ transformation.

For the Weyl multiplet, we established that the vielbein and the
gravitino transformations are also in line with the $4D$
transformations. We have already obtained the correct expressions of
the R-symmetry gauge fields. The transformations of these fields will
lead to rather complicated expressions that include also the
constrained gauge fields. However, the constraints have a different
form in four and in five dimensions which is related to certain field
redefinitions, and this must be taken into account when comparing. As
we have mentioned above, we have already identified the $4D$ auxiliary
tensor field, and likewise we can deduce the expressions for the $5D$
fields $\chi^i$ and $D$ from the explicit variations in terms of the
$4D$ fields. The corresponding formulae will be presented in
\eqref{eq:phi-chi-D}. Some of these results are convenient when
comparing $4D$ and $5D$ actions related by dimensional reduction.

The final result of this section is that the off-shell dimensional
reduction of $5D$ multiplets in a superconformal background can be
carried out systematically. The transformations originating from the
$5D$ multiplets are identical to those in $4D$ up to field-dependent
S-supersymmetry and R-symmetry transformations. This makes the actual
identification of the proper $4D$ fields non-trivial. The
resulting $4D$ theory can understood as a gauge-fixed version of the
standard theory. The gauge-fixing is related to the extra R-symmetry 
that arises in lower dimensions. Both these features are generic.

\section{Off-shell dimensional reduction; matter multiplets}
\label{sec:shell-dimens-reduct-matter}
\setcounter{equation}{0}
In this section we repeat the same analysis as in the previous
section, but now applied to the vector multiplet and the
hypermultiplet. We refrain from presenting similar results for tensor
multiplets. They can be derived by the same method, or, alternatively,
they can be found by considering a composite tensor multiplet
constructed from the square of a vector multiplet.

In five space-time dimensions the vector supermultiplet consists of a
real scalar $\sigma$, a gauge field $W_\mu$, a triplet of (auxiliary)
fields $Y^{ij}$, and a fermion field $\Omega^i$. Under Q- and
S-supersymmetry these fields transform as follows,
\begin{align}
  \label{eq:sc-vector-multiplet}
  \delta \sigma =&\,
  \mathrm{i}\bar{\epsilon}_i\Omega^i \,,
  \nonumber \\ 
  \delta\Omega^i =&\,
  - \ft12 (\hat{F}_{AB}- 4\,\sigma T_{AB}) \gamma^{AB} \epsilon^i
  -\mathrm{i} \Slash{D} \sigma\epsilon^i -2\varepsilon_{jk}\,
  Y^{ij} \epsilon^k 
  + \sigma\,\eta^i \,,   \nonumber\\ 
  \delta W_M =&\,
  \bar{\epsilon}_i\gamma_M\Omega^i - \mathrm{i}
  \sigma \,\bar\epsilon_i \psi_M{}^i  \,, \nonumber\\ 
  \delta Y^{ij}=&\,  
   \varepsilon^{k(i}\, \bar{\epsilon}_k \Slash{D} \Omega^{j)} 
  + 2{\mathrm{i}}\varepsilon^{k(i}\, \bar\epsilon_k (-\ft1{4}
  T_{AB}\gamma^{AB}\Omega^{j)}+ 4 \sigma \chi^{j)})
  -\ft1{2}{\mathrm{i}}  \varepsilon^{k(i}\, \bar{\eta}_k
  \Omega^{j)} \,.  
\end{align}
where $(Y^{ij})^\ast\equiv Y_{ij}= \varepsilon_{ik}\varepsilon_{jl}
Y^{kl}$, and the supercovariant field strength is defined as, 
\begin{equation}
  \label{eq:W-field-strength}
 \hat F_{MN}(W) = 2\, \partial_{[M} W_{N]}  -
 \bar\Omega_i\gamma_{[M} \psi_{N]}{}^i +\ft12
 \mathrm{i}\sigma\,\bar\psi_{[M i} \psi_{N]}{}^i \,.
\end{equation}
The fields behave under local scale transformations according to the
weights shown in table~\ref{tab:w-weights-matter-5D}. 

%
\begin{table}[t]
\centering
\begin{tabular}{|c|cccc|cc|} 
\hline 
 &  \multicolumn{4}{c|}{vector multiplet}&\multicolumn{2}{c|} {hypermultiplet} \\ \hline \hline 
 field & $\sigma$ & $W_\mu$ & $\Omega_i$  & $Y_{ij}$ &$A_i{}^\alpha$&
 $\zeta^\alpha$  \\[.4mm] \hline 
$w$  & $1$ & $0$ &$\tfrac{3}{2}$ & 2 & $\tfrac{3}{2}$ & $2$ \\[.4mm] \hline

\end{tabular}
\vskip 2mm
\renewcommand{\baselinestretch}{1}
\parbox[c]{7.4cm}{\caption{\footnotesize Weyl weights $w$ of the
    vector multiplet and the hypermultiplet component fields in five
    space-time dimensions. \label{tab:w-weights-matter-5D}}}   
\end{table}

The reduction proceeds in the same way as before, except that we have
now the advantage that we have already identified some of the $4D$
fields belonging to the $4D$ Weyl multiplet. We decompose the $5D$
gauge field $W_M$ into a four-dimensional gauge field $W_\mu$ and a
scalar $W= W_5$ by using the standard Kaluza-Klein ansatz, and write
the Q- and S-transformation rules, including the compensating Lorentz
transformation \eqref{eq:comp-Lor}. Just as in
\eqref{eq:compensating-chiral-tr} we introduce an R-covariant spinor
field field $\Omega^i$ by
\begin{equation}
  \label{eq:new-Omega}
  (\Omega^i-\phi^2 W\,\psi^i)\big|^\mathrm{Rcov}= \exp[-\tfrac12\mathrm{i}
  \varphi\,\gamma^5]\, (\Omega^i-\phi^2 W\,\psi^i)\,,
\end{equation}
which transforms under $\mathrm{U}(1)$. In terms of the R-covariant spinor
fields, we derive the following transformation rules,
\begin{align}
  \label{eq:vector}
  \delta\big[\mathrm{e}^{\mp\mathrm{i}\varphi} (\sigma
  \pm\mathrm{i}\phi W)\big] =&\, 2\mathrm{i}
  \bar\epsilon_i \big(\Omega^i-\phi^2W\,\psi^i\big)_\pm\,,
  \nonumber\\[.2ex]  
  \delta W_\mu=&\, \bar\epsilon_i\big[\gamma_\mu(\Omega^i-\phi^2
  W\,\psi^i)_-
  -\mathrm{i} (\sigma-\mathrm{i} \phi
  W)\mathrm{e}^{\mathrm{i}\varphi} 
  \psi_{\mu}{}^i{}_+\big] \nonumber\\
  &\, 
  + \bar\epsilon_i\big[\gamma_\mu(\Omega^i-\phi^2
  W\,\psi^i)_+
  -\mathrm{i} (\sigma+ \mathrm{i} \phi
  W)\mathrm{e}^{-\mathrm{i}\varphi} \psi_{\mu}{}^i{}_-\big]
\,,  \nonumber\\[.2ex] 
  \delta\big(\Omega^i-\phi^2W\,\psi^i\big)_\pm =&\, -\tfrac12
  \big[\hat F(W)_{ab}-\tfrac18(\sigma\mp\mathrm{i} \phi W) \,\hat
  T_{ab} \big] \gamma^{ab}\epsilon^i_\pm \nonumber\\ 
&\,
  - \mathrm{i} \hat{\Slash{D}}
  \big[(\sigma \pm\mathrm{i}\phi
  W)\mathrm{e}^{\mp\mathrm{i}\varphi}\big] \epsilon^i_\mp
  -2 \hat Y^{ik}\varepsilon_{kj} \,\epsilon^j_\pm 
  \nonumber\\
  &\, + (\sigma\pm\mathrm{i} \phi W)\mathrm{e}^{\mp\mathrm{i}\varphi}
  \,\eta^i_\pm  \,,
\end{align}
where $\hat Y^{ij}$ is defined by 
\begin{equation}
  \label{eq:hat-Y}
  \hat Y^{ij} = Y^{ij} + \tfrac12  W\,\hat{V}_k{}^{i} \, \varepsilon^{jk} +
  \tfrac12\phi\,(\bar\Omega_k\gamma^5-\tfrac{1}{2}\mathrm{i}\,\sigma
  \phi\, \bar\psi_k)\psi^{(i} \, \varepsilon^{j)k} \,.
\end{equation} 
Note that in \eqref{eq:vector}, we have again
suppressed the field-dependent S-supersymmetry and $\mathrm{SU}(2)$
R-symmetry transformations.
 
Hypermultiplets are associated with target spaces of dimension $4r$
that are hyperk\"ahler cones \cite{deWit:1999fp}. The supersymmetry
transformations are most conveniently written in terms of the sections
$A_i{}^\alpha(\phi)$, where $\alpha= 1,2,\ldots,2r$,
\begin{align} 
  \label{eq:hypertransf}
  \delta A_i{}^\alpha=&\, 2\mathrm{i}\,\bar\epsilon_i\zeta^\alpha\,,
  \nonumber\\ 
  \delta\zeta^\alpha =&\, - \mathrm{i}\Slash{D}
  A_i{}^\alpha\epsilon^i 
  + \tfrac3{2} A_i{}^\alpha\eta^i \,.
\end{align}
The $A_i{}^\alpha$ are the local sections of an
$\mathrm{Sp}(r)\times\mathrm{Sp}(1)$ bundle. We also note the
existence of a covariantly constant skew-symmetric tensor
$\Omega_{\alpha\beta}$ (and its complex conjugate
$\Omega^{\alpha\beta}$ satisfying
$\Omega_{\alpha\gamma}\Omega^{\beta\gamma}= -\delta_\alpha{}^\beta$),
and the symplectic Majorana condition for the spinors reads as
$C^{-1}\bar\zeta_\alpha{}^\mathrm{T} = \Omega_{\alpha\beta}
\,\zeta^\beta$. Covariant derivatives contain the $\mathrm{Sp}(r)$
connection $\Gamma_A{}^\alpha{}_\beta$, associated with rotations of
the fermions. The sections $A_i{}^\alpha$ are pseudo-real, i.e. they
are subject to the constraint, $A_i{}^\alpha \varepsilon^{ij}
\Omega_{\alpha\beta} = A^j{}_\beta\equiv (A_j{}^\beta)^\ast$. The
information on the target-space metric is contained in the so-called
hyperk\"ahler potential. For our purpose the geometry of the
hyperk\"ahler cone is not relevant. Hence we assume that the cone is
flat, so that the target-space connections and curvatures will
vanish. The extension to non-trivial hyperk\"ahler cone geometries is
straightforward.

For the local scale transformations we refer again to the weights
shown in table~\ref{tab:w-weights-matter-5D}. The hypermultiplet is
not realized as an off-shell supermultiplet. Closure of the
superconformal transformations is only realized upon using fermionic
field equations, but this fact does not represent a serious problem in
what follows. The $4D$ fields have, however, different Weyl weights as
is shown in table \ref{table:w-weights-matter-4D}. This has been taken
into account in the reduction, by scaling $A_i{}^\alpha$ by a factor
$\phi^{-1/2}$, as can be seen below. Furthermore we define an
R-covariant spinor combination,
\begin{equation}
  \label{eq:zeta-new}
  (\phi^{-1/2} \zeta^\alpha-\tfrac12\phi^{1/2} A_j{}^\alpha \gamma^5
  \psi^j)\big|^\mathrm{Rcov}= \exp[\tfrac12\mathrm{i} 
  \varphi\,\gamma^5]\, (\phi^{-1/2} \zeta^\alpha-\tfrac12\phi^{1/2}
  A_j{}^\alpha \gamma^5 \psi^j)\,. 
\end{equation}

The $5D$ Q- and S-supersymmetry variations take the following
form, again in terms of R-covariant chiral spinors, 
\begin{align}
  \label{eq:hyper}
  \delta( \phi^{-1/2} A_i{}^\alpha) =&\,
  2 \mathrm{i}\,\bar\epsilon_i\big(\phi^{-1/2} \zeta^\alpha
  -\tfrac{1}2\mathrm{i}\, \phi^{1/2} A_j{}^\alpha \gamma^5
  \psi^j\big)_+ \nonumber\\
  &\,+ 2 \mathrm{i}\,\bar\epsilon_i\big(\phi^{-1/2} \zeta^\alpha
  -\tfrac{1}2\mathrm{i}\, \phi^{1/2} A_j{}^\alpha \gamma^5
  \psi^j\big)_- \,,   \nonumber\\
  \delta \big(\phi^{-1/2} \zeta^\alpha -\tfrac{1}2\mathrm{i}\,
  \phi^{1/2} A_j{}^\alpha \gamma^5 \psi^j\big)_\pm =&\,
  - \mathrm{i}\Slash{\hat D}( \phi^{-1/2}
  A_i{}^\alpha)\,\epsilon^i_\mp  +\,\phi^{-1/2} A_i{}^\alpha
  \eta^i_\pm  \,,
\end{align}
where, as before, we suppressed the S-supersymmetry and R-symmetry
transformations with field-dependent parameters as specified by
\eqref{eq:field-dep-S-R}. Note that the proportionality factor in
front of the $4D$ S-supersymmetry variation has changed as compared to
the $5D$ result \eqref{eq:hypertransf}. 

%
\begin{table}[t]
\begin{center}
\begin{tabular*}{10.0cm}{@{\extracolsep{\fill}}|c|cccc|cc| }
\hline
 & \multicolumn{4}{c|}{vector multiplet} & 
 \multicolumn{2}{c|}{hypermultiplet} \\
 \hline \hline
 field & $X$ & $W_\mu$  & $\Omega_i$ & $Y^{ij}$& $A_i{}^\alpha$ & $\zeta^\alpha$ \\[.5mm] \hline
$w$  & $1$ & $0$ & $\tfrac32$ & $2$ & $1$ &$\tfrac32$
\\[.5mm] \hline
$c$  & $-1$ & $0$ & $-\tfrac12$ & $0$ & $0$ &$-\tfrac12$
\\[.5mm] \hline
$\gamma_5$   & && $+$  &   &  &$-$ \\ \hline
\end{tabular*}
\vskip 2mm
\renewcommand{\baselinestretch}{1}
\parbox[c]{10.0cm}{\caption{\footnotesize Weyl
      and chiral weights ($w$ and $c$) and fermion chirality
      $(\gamma_5)$ of the vector multiplet and the hypermultiplet
      component fields in four space-time
      dimensions. \label{table:w-weights-matter-4D}  }}
\end{center}
\end{table}

\section{Five and four-dimensional fields and invariant Lagrangians}
\label{sec:five-four-dimens-fields-lagr}
\setcounter{equation}{0}
After expressing the $5D$ spinors into chiral $4D$ spinors according
to the procedure explained in appendix \ref{sec:spinors-four-five}, we
can identify the $4D$ fields in terms of the $5D$ ones. Note that for
matter fields the overall normalization of the components is only
determined up to a real constant. For the vector multiplet we choose
the normalization such that the vector gauge field remains the
same. This has the advantage that we can easily compare the
corresponding charges in four and five dimensions. Phase factors can
be changed according to the chiral $\mathrm{U}(1)$ transformations
which constitute an invariance of the theory, but they should be
applied uniformly. Our four-dimensional transformations coincide with
those given in \cite{deWit:2010za}. For the convenience of the reader
we have also included a summary in appendix \ref{App:SC}.

We thus express the $4D$ fields in terms of the $5D$ fields and the
field $\varphi$ for each multiplet separately. First we present the
Kaluza-Klein and the matter vector multiplets, then the
hypermultiplet, and finally we turn to the Weyl multiplet.\\
{\it The Kaluza-Klein vector multiplet:}
\begin{align}
  \label{eq:KK-identification}
  X^0=&\, -\tfrac12 \phi\,\mathrm{e}^{-\mathrm{i}\varphi} \,,\nonumber\\
  \Omega_i{}^0 =&\, -\varepsilon_{ij}\,\phi^2 \,
  \mathrm{e}^{-\tfrac12\mathrm{i}\varphi} \,\psi^j_+ \,,\qquad 
    \Omega^{i\,0} =\mathrm{i}
    \phi^2\,\mathrm{e}^{\tfrac12\mathrm{i}\varphi}\,\psi^i_-
    \,,\nonumber\\ 
    W_\mu{}^0=&\, B_\mu\,,\nonumber\\
    Y^{ij\, 0} =&\,  \hat V_k{}^i\,\varepsilon^{jk}  \,,
\end{align}
{\it The matter vector multiplet:}
\begin{align}
  \label{eq:matter-vector-identification}
  X=&\, -\tfrac12 \mathrm{i} (\sigma+\mathrm{i}\phi W)\,
  \mathrm{e}^{-\mathrm{i}\varphi} \,,\nonumber\\ 
  \Omega_i =&\,
  -\varepsilon_{ij}\,\mathrm{e}^{-\tfrac12\mathrm{i}\varphi}(\Omega^j -
  \phi^2W\psi^j)_+ \,,\qquad 
    \Omega^i = \mathrm{i}
    \mathrm{e}^{\tfrac12\mathrm{i}\varphi}\,(\Omega^i
    -\phi^2W\psi^i)_- \,,\nonumber\\  
    W_\mu=&\, W_\mu\,,\nonumber\\
    Y^{ij} =&\,-2\,\hat Y^{ij}\,,
\end{align}
{\it The hypermultiplet:}
\begin{align}
  \label{eq:matter-hyper-identification}
  A_i{}^\alpha =&\, \phi^{-1/2} A_i{}^\alpha \,,\nonumber\\ 
  \zeta^\alpha =&\,\mathrm{e}^{-\tfrac12\mathrm{i} 
  \varphi} \, (\phi^{-1/2} \zeta^\alpha-\tfrac12\phi^{1/2}
  A_j{}^\alpha \gamma^5 \psi^j)_- \,,\nonumber\\ 
    \zeta_\alpha =&\,-\mathrm{i} \Omega_{\alpha\beta}\,
    \mathrm{e}^{\tfrac12\mathrm{i}  \varphi}\, 
    (\phi^{-1/2} \zeta^\beta-\tfrac12\phi^{1/2}  A_j{}^\beta \gamma^5
    \psi^j)_+\,. 
\end{align}
{\it The Weyl multiplet:}
\begin{align}
  \label{eq:Weyl-Weyl-identification}
  e_\mu{}^a=&\, e_\mu{}^a \,,\nonumber\\
  \psi_\mu{}^i =&\, \mathrm{e}^{-\tfrac12\mathrm{i}\varphi}
  \,\psi_\mu{}^i_+ \,, \qquad
  \psi_{\mu i} = \mathrm{i} \varepsilon_{ij}\,
  \mathrm{e}^{\tfrac12\mathrm{i}\varphi} \,\psi_\mu{}^j_- \,,\nonumber\\
  T_{ab}{}^{ij} =&\,-\tfrac12\mathrm{i}
  \mathrm{e}^{-\mathrm{i}\varphi} \,\hat T_{ab}^-\,\varepsilon^{ij}   \,,\nonumber\\
  \mathcal{V}_\mu{}^{i}{}_j =&\, \hat V_{\mu j}{}^i \,,\qquad 
  b_\mu=\, b_\mu\,, \qquad A_\mu =\, \hat A_\mu
  + \partial_\mu\varphi\,. 
\end{align}
The remaining fermion fields of the $5D$ Weyl multiplet, $\phi_M{}^i$
and $\chi^i$, follow from the $5D$ Q-supersymmetry transformations of
$b_M$ and $ V_{Mi}{}^j$.  Likewise the remaining fermions of the $4D$
Weyl multiplet, $\phi_\mu{}^i$ and $\chi^i$, follow from the
Q-supersymmetry variations of $b_\mu$ or $\mathcal{V}_{\mu j}{}^i$,
and $A_\mu$. To disentangle the two sets of fermion fields one makes
use of the conventional constraints. The relevant $5D$ constraint was
given in \eqref{eq:conv-constraints-5} and the corresponding $4D$
constraints are given in \eqref{eq:conv-constraints}. The same comment
applies to the composite gauge fields $f_M{}^A$ and $f_\mu{}^a$
corresponding to the $5D$ and $4D$ conformal boosts,
respectively. Finally one determines the scalar field $D$ from
considering the variation of the field $\chi^i$. We summarize some of
the relevant results below, suppressing terms of higher order in the
fermion fields,
\begin{align}
  \label{eq:phi-chi-D}
  f_a{}^a\vert_{4D}=&\, f_a{}^a\vert_{5D} - \tfrac12\,D\vert_{4D}
  -\tfrac1{16} \phi^{-2} \, F(B)^{ab}F(B)_{ab}\,, \nonumber \\[.2ex]
   \phi_{\mu}{}^{i}\big\vert_{4D} =&\,
   2 \mathrm{i} \phi_{\mu}{}^{i}\big\vert_{5D} +4\gamma_\mu\chi^{i}
   +(e^a_\mu-\tfrac38\,\gamma_\mu\gamma^a) \gamma_5 \mathcal{D}_a \psi^{i}
   + \tfrac3{32} F(B)^{ab}\big(\gamma_\mu\gamma_{ab}
   -2\,\gamma_{ab} \gamma_\mu \big) \psi^{i}\,, \nonumber\\
   & -\tfrac38\mathrm{i}\phi  T^{ab} \gamma^5\big(\gamma_\mu\gamma_{ab}-
   2\,\gamma_{ab} \gamma_\mu \big) \psi^{i}
   -\tfrac3{8}\mathrm{i}\phi T_{a5}\big( 10\,e_\mu{}^a - \gamma^a\gamma_\mu
   \big) \psi^{i}   \nonumber\\
   &-\tfrac12 \left(\Slash{\mathcal{D}} \phi \, \gamma_\mu
     +\tfrac32\,\gamma_\mu \Slash{\mathcal{D}} \phi \right)\gamma^5 \psi^{i}
   -\tfrac14 \phi^2 \mathcal{V}^{i}{}_j \gamma_\mu \psi^j\,,
   \nonumber\\[.2ex]
   \chi^i \big\vert_{4D}   =&\,
   8 \chi^i +\tfrac{1}{48} \gamma^{ab}F(B)_{ab}\psi^i
   -\tfrac3{4}\mathrm{i} \,\phi\, T_{ab} \gamma^5 \gamma^{ab}\psi^i \,,
   \nonumber\\
   &+\tfrac1{4} \phi^{-1} \,\gamma_5 \Slash{\mathcal{D}}( \phi^2\psi^i)
   - \tfrac12 \phi^{2} \mathcal{V}^i{}_j \psi^j
   - \tfrac94\mathrm{i}  \phi\, T_{a5}\gamma^a \psi^i \,,\nonumber
   \\[.2ex]
  D\big\vert_{4D}  =&\,4\, D\big\vert_{5D} + \tfrac1{4} \phi^{-1}\big(
  \mathcal{D}^a\mathcal{D}_a +\tfrac16R\big)\phi +
  \tfrac3{32}\phi^{-2} \, F(B)^{ab}F(B)_{ab}\nonumber\\
  &\,+ \tfrac32 T^{ab}T_{ab} +3\, T^{a5}T_{a5} +\tfrac1{4}
  \phi^2 \, V_i{}^j\, V_j{}^i \,,
\end{align}
where $D^aD_a\phi = (\mathcal{D}^a\mathcal{D}_a + \tfrac16 R) \phi$
equals the $4D$ conformally invariant D'Alembertian with $R$ the $4D$
Ricci scalar.  One can proceed and rewrite the covariant derivatives
on the spinors in terms of $4D$ fields, to verify that the
supersymmetry variations of the fields above are indeed identical to
the ones in four dimensions, but this is not necessary here. The only
result we will need in the remainder of this section is the last
expression for the field $D$. 

Suppressing the higher-order fermionic contributions we now express
the $5D$ bosonic fields into the $4D$ ones.  We assume that $\phi$ is positive
so that we are considering compactification of a space-like
coordinate. The $5D$ components of the metric are already specified
in \eqref{eq:kk-ansatz}. The remaining expressions are,
\begin{align}
 \label{eq:inverse-5-to-4}
 \phi =&\, 2\, \vert X^0\vert\,,\nonumber\\[.4ex]
 B_\mu=&\, W_\mu{}^0\,, \nonumber\\[.4ex]
 V_{M i}{}^j =&\, \left\{ \begin{array}{rl}
 V_{\mu i}{}^j =&\!\! \mathcal{V}_\mu{}^j{}_i - \tfrac14
 \varepsilon_{ik}\, Y^{kj\,0}\,\vert X^0\vert^{-2} \, W_\mu{}^0\,, \\[.6ex]
 V_{5i}{}^j  =&\!\!-\tfrac14\varepsilon_{ik}\, Y^{kj\,0}\,\vert
 X^0\vert^{-2}   \,,
  \end{array} \right.   \nonumber\\[.4ex]
 \sigma=&\, -\mathrm{i} \vert X^0\vert\,(t-\bar t)\,, \nonumber\\[.4ex]
 W_M=&\,\left\{ \begin{array}{rl}
     W_\mu=& \!\!W_\mu -\tfrac12 (t+\bar t) \,W_\mu{}^0 \,, \\[.6ex]
 W_5 =& \!\!-\tfrac12(t+\bar t)\,,
  \end{array} \right.   \nonumber\\[.4ex]
Y^{ij}=&\,-\tfrac12 Y^{ij} +\tfrac14 (t+\bar t)\,Y^{ij\,0} \,,
\nonumber\\[.4ex]
T_{a5} =&\, \tfrac1{12} \mathrm{i} \,e_a{}^\mu \Big(
\frac{\mathcal{D}_\mu X^0}{X^0} -
\frac{\mathcal{D}_\mu \bar X^0}{\bar X^0}\Big)\,,\nonumber\\[.4ex]
T_{ab} =&\, - \frac{\mathrm{i}}{24\,\vert X^0\vert}
\Big(\varepsilon_{ij}T_{ab}{}^{ij} \,\bar X^0 - F_{ab}^-{}^0 \Big)
+\mathrm{h.c.}  \,,
\end{align}
where $\mathcal{D}_\mu X^0=(\partial_\mu-b_\mu +\mathrm{i}A_\mu)X^0$
and $t=X/X^0$, and all the fields on the right-hand side refer to $4D$
fields. 
 
In the remainder of this section we evaluate the reduction of $5D$
supersymmetric actions to four dimensions. We concentrate on actions
for vector multiplets and for hypermultiplets, both at most quadratic
in derivatives, and on a third action that contains terms proportional
to the square of the Riemann tensor accompanied by other terms quartic in
derivatives. In four dimensions, four different invariant actions are
expected to be generated, related to the fact that there exists a
second class of actions with higher-derivative couplings associated
with the vector multiplets (see, e.g. \cite{deWit:2010za} and
references quoted therein). However, what we will establish below, is
that there exists yet another higher-derivative action that involves
terms quadratic in the Ricci tensor. This action has not appeared in
the literature so far.

The $5D$ bosonic Lagrangian for hypermultiplets reads
\begin{equation}
  \label{eq:5D-hyper-lagr}
  8\pi\,\mathcal{L}_\mathrm{hyper} = -\tfrac12 E\,
  \Omega_{\alpha\beta}\, \varepsilon^{ij}\big\{ 
  {\cal D}_M A_i{}^\alpha\, {\cal D}^M
  A_j{}^{\beta}- A_i{}^\alpha A_j{}^\beta \big[\tfrac{3}{16} R
  + 2\, D  + \tfrac3{4} T^{AB}T_{AB} \big] \Big\} \,.
\end{equation} 
Upon reduction to four dimensions, the first term becomes 
\begin{align}
  \label{eq:DADA-terms}
  -\tfrac12 E\,\Omega_{\alpha\beta}\, \varepsilon^{ij} \, {\cal D}_M
  A_i{}^\alpha\, {\cal D}^M A_j{}^{\beta} =&\,  -\tfrac12
  e\,\Omega_{\alpha\beta}\, \varepsilon^{ij} \big\{ {\cal D}_M 
  A_i{}^\alpha\, {\cal D}^M A_j{}^{\beta}  \\
  &\, +  A_i{}^\alpha A_j{}^\beta
  \big[-\tfrac12 \mathcal{D}^\mu[\phi^{-1}\mathcal{D}_\mu \phi]
  +\tfrac14\phi^{-2} [\mathcal{D}_\mu\phi]^2 
  -\tfrac18 \phi^2\, V_k{}^l V_l{}^k \big]\big\} \,,\nonumber
\end{align}
where we suppressed a total derivative. Next we turn to the second
term in \eqref{eq:5D-hyper-lagr}. Making use of
\eqref{eq:contracted-R}, which relates the $5D$ and $4D$ Ricci
scalars, and the relation between the $4D$ and $5D$ $D$-fields, the
combination of the two terms readily combines into
\begin{align} 
  8\pi^2\,e^{-1}\mathcal{L}_\mathrm{hyper} =&\,
  -\tfrac12\,\phi^{-1} \Omega_{\alpha\beta}\, \varepsilon^{ij} \Big\{
  {\cal D}_\mu A_i{}^\alpha\, {\cal D}^\mu A_j{}^{\beta}
   - A_i{}^\alpha A_j{}^\beta \big( \tfrac12 D  + \tfrac16 R\big)
   \Big\} \,,
\end{align}
which agrees with the well-known expression for the supersymmetric
$4D$ Lagrangian \cite{deWit:1984px}. Observe that the Kaluza-Klein
vector multiplet decouples from the hypermultiplets, as it should.

Subsequently we turn to the $5D$ bosonic Lagrangian for vector
multiplets,whose evaluation is somewhat more cumbersome,  
\begin{align}
  \label{eq:vector-5D-Lagr} 
   8\pi^2\mathcal{L}_\mathrm{vvv} =&\,
  3\, E\,C_{ABC} \,\sigma^A\Big[\tfrac12 \mathcal{D}_M\sigma^B
  \,\mathcal{D}^M\sigma^C
  + \tfrac14 F_{MN}{}^B F^{MN C} - Y_{ij}{}^B  Y^{ijC }-3\,\sigma^B
  F_{MN}{}^C  T^{MN}\Big] \nonumber \\
  &\,
  - \tfrac1{8}\mathrm{i}\,C_{ABC}  \varepsilon^{MNPPQR}\, W_M{}^A
  F_{NP}{}^B F_{QR}{}^C
  \nonumber \\
  &\, 
  - E\,C_{ABC}\sigma^A\sigma^B\sigma^C\,  \Big[\tfrac18
  R - 4\,D - \tfrac{39}2 T^{AB}T_{AB}\Big]\,,
\end{align}

The first term is rewritten as,
\begin{align}
  \label{eq:first-term-vector}
  &3\,E\,C_{ABC} \,\sigma^A\Big[\tfrac12 \mathcal{D}_M\sigma^B
  \,\mathcal{D}^M\sigma^C + \tfrac14 F_{MN}{}^B F^{MN C} - Y_{ij}{}^B
  Y^{ijC }-3\,\sigma^B
  F_{MN}{}^C  T^{MN}\Big] = \nonumber \\
  &\qquad -3\,\mathrm{i} e \vert X^0\vert^2 \, C_{ABC} (t-\bar t)^A
  \,\mathcal{D}^\mu t^B \,\mathcal{D}_\mu \bar t^C
  \nonumber\\
  &\qquad + \tfrac34\mathrm{i} e\, C_{ABC} (t-\bar t)^A (t-\bar t)^B (t-\bar
  t)^C \, (\mathcal{D}_\mu\vert X^0\vert)^2 
  \nonumber\\
  &\qquad -\tfrac32\mathrm{i} e\, C_{ABC} (t-\bar t)^A (t-\bar t)^B
  \big(\bar X^0\,\mathcal{D}^\mu \bar t^C \,\mathcal{D}_\mu X^0-
  X^0\,\mathcal{D}^\mu t^C\,\mathcal{D}_\mu\bar X^0  \big) \nonumber\\
  &\qquad -\tfrac3{8}\mathrm{i}e\, C_{ABC} (t-\bar t)^A \Big[ F_{ab}{}^B
  \,F^{ab C} - F_{ab}{}^B\, F^{ab 0} (t+\bar t)^C + \tfrac14 \,
  (F_{ab}{}^0)^2\, (t+\bar t)^B(t+\bar t)^C \Big]\nonumber\\
  &\qquad +\tfrac3{8}\mathrm{i}e\, C_{ABC} (t-\bar t)^A \Big[ Y_{ij}{}^B
  Y^{ij C} - (t+\bar t)^B Y_{ij}{}^C Y^{ij 0} +\tfrac14 (t+\bar
  t)^B(t+\bar t)^C\,  \vert Y_{ij}{}^0\vert^2 \Big] \nonumber \\
  &\qquad -\tfrac3{16}\mathrm{i} e\,C_{ABC} (t-\bar t)^A (t-\bar t)^B
  \Big[\big(F_{ab}{}^C -\tfrac12(t+\bar t)^C F_{ab}{}^0\big)\, \big(
  \varepsilon_{ij} T^{ab ij}\bar X^0  - F^{-ab 0} \big) - \mathrm{h.c.}
  \Big] \,,
\end{align}
where we employed special coordinates $t^A= X^A/X^0$. The $5D$
Chern-Simons term can be rewritten as follows,
\begin{align}
  \label{eq:CS-term}
   &- \tfrac1{8}\mathrm{i}\,C_{ABC}\,
  \varepsilon^{MNPQR} W_M{}^A
  F_{NP}{}^B F_{QR}{}^C = \tfrac1{64} \mathrm{i}\, C_{ABC}\,
  \varepsilon^{\mu\nu\rho\sigma} \Big[ 12(t+\bar t)^A
  F_{\mu\nu}{}^BF_{\rho\sigma}{}^C \nonumber\\
  &\qquad  -6(t+\bar t)^A(t+\bar t)^B
  F_{\mu\nu}{}^CF_{\rho\sigma}{}^0 + (t+\bar t)^A(t+\bar t)^B(t+\bar
  t)^C F_{\mu\nu}{}^0 F_{\rho\sigma}{}^0 \Big] \,,
\end{align}
and finally the last term is rewritten as, 
\begin{align}
  \label{eq:third-line-vector-Lagr}
    &- E\,C_{ABC}\sigma^A\sigma^B\sigma^C\,  \Big[\tfrac18
  R - 4\,D - \tfrac{39}2 T^{AB}T_{AB}\Big]= -\tfrac1{2}\mathrm{i} e\,
  C_{ABC} (t-\bar t)^A (t-\bar t)^B (t-\bar t)^C  \nonumber\\
  &\quad \times \Big[(\tfrac16 R- D)\,\vert X^0\vert^2  - \tfrac1{16}
  Y_{ij}{}^0 \,Y^{ij0} -\vert\mathcal{D}_\mu X^0\vert^2 +\tfrac32
  \big(\mathcal{D}_\mu\vert X^0\vert \big)^2 +\tfrac1{32} F_{ab}{}^0
  F^{ab 0} \nonumber\\
   &\qquad 
  + \tfrac1{32} \big[\big(
   \varepsilon_{ij} T_{ab}{}^{ij}\bar X^0 - F_{ab}^-{}^0\big)^2  +
   \mathrm{h.c.}\big] \Big] \,. 
\end{align}

The resulting Lagrangian can be expressed in terms of the following
homogeneous and holomorphic function of degree two \cite{deWit:1984pk},
\begin{equation}
  \label{eq:functio-F}
  F(X) =- \frac12 \frac{C_{ABC} X^AX^BX^C}{X^0}\,, 
\end{equation}
which encodes the bosonic terms of the Lagrangian according to
\begin{align}
\label{eq:4D-lagrangian-vectors}
  e^{-1} \mathcal{L}_\mathrm{bosonic} =&\, -\mathrm{i}
  \big(\mathcal{D}_\mu X^I \,\mathcal{D}^\mu \bar F_I -\mathcal{D}_\mu
  \bar X^I  \,\mathcal{D}^\mu F^I \big) +\mathrm{i} (X^I\bar F_I -
  \bar X^I F_I) \big(\tfrac{1}{6} R -D\big) \nonumber\\
  &\,+ \ft14
  \mathrm{i}\left[F_{IJ} \,F^{-\,I}_{\mu\nu}
  F^{-\,\mu\nu J} -\bar F_{IJ} \,F^{+\,I}_{\mu\nu}  F^{+\mu\nu
    J}\right] \nonumber\\[.6ex] 
  &\,+\big[\tfrac1{8}
  \bar X^I  N_{IJ} F^{-ab I} \, T_{ab}{}^{ij}\varepsilon_{ij}
  -\tfrac1{64} \bar X^I  N_{IJ}  \bar X^J 
  \, \big(T_{\mu\nu}{}^{ij}\varepsilon_{ij}\big)^2 +\mbox{h.c.}
  \big]\nonumber\\
  &\,
    + N_{IJ}\,Y_{ij}{}^I Y^{ij J}  \,. 
\end{align}
where $N_{IJ}= -\mathrm{i} F_{IJ} + \mathrm{i} \bar F_{IJ}$

Finally, we turn to the reduction of the four-derivative coupling
involving the vector multiplets and the Weyl multiplet, first
introduced in \cite{Hanaki:2006pj}. Here, we refrain from giving full
details of the invariants in both five and four dimensions,
concentrating on the identification of the relevant functions arising
under dimensional reduction. We use the conventions of
\cite{deWit:2009de} and concentrate on the following
terms\footnote{ 
  Here we use the index $A$ to label the $5D$ vector multiplets and
  the indices $B,C,\ldots$ to indicate $5D$ tangent-space
  indices.}, 
\begin{align}
  \label{eq:vww}
  8\pi^2\,\mathcal{L}_\mathrm{vww} =&{}\,
  \tfrac1{4}\,E\, c_A Y_{ij}{}^A \, T^{CD} R_{CD k}{}^j(V) \,\varepsilon^{ki}
  \nonumber \\[.5ex]
  &{}
  +E\, c_A\sigma^A\,\Big[
  \tfrac1{64} R_{CD}{}^{EF}(M)\,R_{EF}{}^{CD}(M) +\tfrac1{96}
  R_{MN j}{}^i(V) \,R^{MN}{}_i{}^j(V) \Big]  \nonumber\\[.5ex]
  &{}
  -\tfrac1{128}\mathrm{i} \varepsilon^{MNPQR}\,c_AW_M{}^A\,\left[
    R_{NP}{}^{CD}(M)\,R_{QRCD}(M)
    + \tfrac13\, R_{NP j}{}^i(V) \,R_{QR i}{}^j(V)\right]
  \nonumber\\[.5ex] 
  &\, +\tfrac3{16} E\,  c_A\big(10\,\sigma^A T_{BC}-F_{BC}{}^A\big)
  R(M)_{DE}{}^{BC}\, T^{DE} +\cdots\,, 
\end{align} 
where $R(M)_{MN}{}^{CD}$ coincides with the $5D$ Weyl tensor, up
to certain additions implied by supersymmetry.  Upon reduction of
\eqref{eq:vww} to four dimensions one obtains a $4D$ supersymmetric
Lagrangian with higher-derivative couplings. For our purpose, it
suffices to concentrate on the terms that involve the tensors
$R(M)_{ab}{}^{cd}$ and/or $R(\mathcal{V})_{ab}{}^i{}_j$. As it turns
out these terms can be decomposed into three sets that exhibit a
mutually different structure. Subsequently we will try and identify
these sets in terms of independent $4D$ supersymmetric Lagrangians.

The first set of terms is given by, 
\begin{align}
  \label{eq:vww1}
  8\pi^2\,\mathcal{L}_\mathrm{vww} \to&\,-\tfrac1{64} \mathrm{i} c_A
  t^A \left[2\, R(M)_{ab}^{-cd}\,R(M)_{cd}^{-ab} +
    R(\mathcal{V})_{ab}^{-i}{}_j \, R(\mathcal{V})^{-abj}{}_i
  \right] \nonumber \\ 
  &\, -\tfrac1{512} \mathrm{i} \varepsilon_{mn}
  T^{abmn} \,(X^0)^{-1} c_A\big(Y^{ijA} - t^A Y^{ij0}\big)\,
  R(\mathcal{V})_{ab}^{-k}{}_j \,\varepsilon_{ki}  \nonumber\\ 
  &\, +\tfrac1{256}\mathrm{i}c_A\, (X^0)^{-1}\varepsilon_{ij} T^{cdij} \, 
  R(M)^{ab}{}_{cd} \big( F^{-A}_{ab}- t^A F^{-0}_{ab} \big)
  \nonumber\\
  &\,  +\mathrm{h.c.} \,, 
\end{align}
which, as we shall see, belongs to a $4D$ supersymmetric invariant
based on a chiral superspace integral
\cite{Bergshoeff:1980is,LopesCardoso:2000qm}. Here $R(M)_{ab}{}^{cd}$
denotes the $4D$ Weyl tensor.

The second set of terms involves expressions that cannot be readily
associated with a known $N=2$ supersymmetric invariant,
\begin{align}
  \label{eq:vww2}
  8\pi^2\,\mathcal{L}_\mathrm{vww} \to&\,- \tfrac1{384}\mathrm{i} c_A
  t^A \left[\tfrac23 \mathcal{R}_{ab}\mathcal{R}^{ab}  +
    R(\mathcal{V})_{ab}^{+i}{}_j \, 
    R(\mathcal{V})^{+abj}{}_i \right] \nonumber\\ 
    &\, -\tfrac1{768}\mathrm{i} c_A(t^A-\bar t^A)\,
    (X^0)^{-1}\varepsilon_{ij} T^{cdij} \,
    R(M)^{ab}{}_{cd}\,F^{-0}_{ab}  \nonumber\\
    &\,  +\mathrm{h.c.}  \,.
\end{align} 
A conspicuous feature of this term is its dependence on the Ricci
tensor $\mathcal{R}_{ab}$. It is rather obvious that this term is
not related to a chiral superspace invariant. The same comment applies
to the third set of terms, given by,
\begin{align}
  \label{eq:vww3}
  8\pi^2\,\mathcal{L}_\mathrm{vww} \to &\,\tfrac1{384}\mathrm{i} c_A \,
  R(\mathcal{V})^{+ab k}{}_j \,\varepsilon_{ki}\, \vert X^0\vert^{-2}
  \big[F^{+A}_{ab} \,Y^{ij0} -F^{+0}_{ab} \,Y^{ijA} + (t-\bar t)^A
  F^{+0}_{ab} \,Y^{ij0} \big]   \nonumber\\
  &\, +\tfrac1{1536} \mathrm{i} \varepsilon^{mn}
  T^{ab}{}_{mn} \,(\bar X^0)^{-1} c_A \big(Y^{ijA} - (2t^A-\bar t^A)
  Y^{ij0}\big)\, 
  R(\mathcal{V})_{ab}^{+k}{}_j \,\varepsilon_{ki}   \nonumber\\ 
  &\,+\mathrm{h.c.} \,. 
\end{align}

We have now completely determined the terms that depend on
$R(\mathcal{V})_{ab}{}^i{}_j$, as well as the terms in \eqref{eq:vww}
that depend explicitly on $R(M)_{ab}{}^{cd}$. However, \eqref{eq:vww}
also contains a term with a double derivative proportional to
$T^{AB}\,D^CD_AT_{BC}$ which can in principle give rise to additional
curvature terms upon reordering derivatives combined with partial
integrations. The evaluation of some of these terms remains therefore
a little ambiguous at this stage, also because the final result may be
subject to the similar rearrangements. Nevertheless the results
determined above are sufficient to discuss the structure of the
resulting $4D$ Lagrangians.

As was mentioned above, the terms \eqref{eq:vww1} exhibit an underlying
holomorphic structure that is characteristic for an invariant based on
a chiral superspace integral (sometimes referred to as an
`F-term'). Such an invariant is well known and it can again be encoded
into a holomorphic function. This function can be included into the
function \eqref{eq:functio-F} by introducing a dependence on an extra
complex field, $\hat A$, which is equal to $\hat
A=(T_{ab}{}^{ij}\varepsilon_{ij})^2$. In the case at hand, the
dependence on $\hat A$ is linear, but for a general $4D$ Lagrangian
the function has to be holomorphic and homogeneous of second degree
\cite{Bergshoeff:1980is,LopesCardoso:2000qm}. As it turns out, the
modified function $F(X,\hat A)$ that correctly encodes the sum of
\eqref{eq:4D-lagrangian-vectors} and \eqref{eq:vww1}, equals,
\begin{equation}
  \label{cub-prepot-ext}
  F(X,\hat A)= -\frac1{2} \frac{C_{ABC}X^A X^B X^C}{X^0}
          -\frac{1}{2048}\,\frac{c_A\,X^A}{X^0}\, \hat A\,,
\end{equation}
where the higher-order derivative Lagrangian encoded in this function
reads as \cite{LopesCardoso:2000qm}, 
\begin{align}
  \label{eq:4d-conf-lagr-R2}
  e^{-1}\, {\cal L}=&\, 
  -4\mathrm{i} \,F_{{\hat A}I}\,T^{cdlm}\varepsilon_{lm} \big[
  2\,R(M)_{cd}{}^{\!ab} \, (F^{-I}_{ab} - \ft14  \bar X^I 
  T_{ab}{}^{ij}\varepsilon_{ij})
  -\varepsilon_{ki}R(\mathcal{V})_{cd}{}^k{}_j \,Y^{ijI} \big]
  \nonumber \\[.5ex] 
  &\,+16 \mathrm{i} F_{\hat A}
  \left[2\, R(M)^{-cd}{}_{\!ab}\,
    R(M)_{ab}^{-cd}  + R({\cal V})^{-ab\,k}{}_l^{~} \,
    R({\cal V})^-_{ab}{}^{\!l}{}_k\right] +\cdots\nonumber \\[.5ex]
  &\, +\mbox{h.c.}\;.
\end{align}
Here we only give the terms relevant for the comparison with
\eqref{eq:vww1}. 

The interpretation of \eqref{eq:vww2} is, however, less clear, as
terms of this type have never been written down explicitly in $N=2$
supergravity. Supersymmetric invariants that contain the square of the
Ricci tensor have been written down in $N=1$ supergravity, often in
the context of a supersymmetrization of the Gauss-Bonnet term
\cite{CERN-TH-2418,LPTENS 78/14,ITP-SB-79-4,PRINT-78-1005
  (KARLSRUHE)}. The latter is a topological invariant whose integral
is proportional to the Euler characteristic of the corresponding
manifold. The emergence of this new supersymmetric coupling in the
reduction from \eqref{eq:vww} constitutes a new result. A brief
perusal of the various terms arising in this reduction shows that it
will involve quite a variety of new couplings. It is obviously of
importance to understand the structure of this invariant and its
possible implications for $4D$ black hole entropy, also in view of the
recent discussion in \cite{arXiv:1108.3842}.

Finally we come to the invariant that contains the terms
\eqref{eq:vww3}. On closer inspection it turns out the the dimensional
reduction of \eqref{eq:vww} involves also terms quartic in the field
strengths. These couplings belong to the class of invariants
constructed in \cite{deWit:2010za}. One of these invariants is indeed
quartic in the field strengths and it contains the following
characteristic terms,
\begin{align}
 \label{eq:real-susp-action}
 e^{-1}\mathcal{L} =&\,
 \tfrac14\,\mathcal{H}_{IJ\bar K \bar L}
   \big( F_{ab}^-{}^I\, F^{-ab\,J}
               -\tfrac12 Y_{ij}{}^I\, Y^{ijJ} \big)
               \big(F_{ab}^+{}^K \, F^{+ab\,L} -\tfrac12 Y^{ijK}\,
                 Y_{ij}{}^L  \big)
             \nonumber\\[.5ex]
  &\,-\Big\{ \mathcal{H}_{IJ\bar K}\big(
  F^{-ab\,I}\, F_{ab}^{-\,J} -\tfrac12 Y^I_{ij}\, Y^{Jij})
     \big( \Box_\mathrm{c} X^K
     + \tfrac18 F^{-\,K}_{ab}\, T^{ab kl}  \varepsilon_{kl}\big)
     +\mathrm{h.c.}\Big\}   \displaybreak[0] \nonumber\\[.5ex]
    &\,+\mathcal{H}_{I\bar J}\Big[ 4\big( \Box_\mathrm{c} \bar X^I + \tfrac18
        F_{ab}^{+\,I}\, T^{ab}{}_{ij} \varepsilon^{ij}\big)
    \big( \Box_\mathrm{c}  X^J + \tfrac18 F_{ab}^{-\,J}\, T^{abij}
      \varepsilon_{ij}\big) \nonumber\\
      & \qquad\qquad +8\,\mathcal{D}_{a}F^{-\,abI\,}\,
      \mathcal{D}_cF^{+c}{}_{b}{}^J   - \mathcal{D}_a Y_{ij}{}^I\,
           \mathcal{D}^a Y^{ij\,J}
    \nonumber\\
    &\qquad\qquad + 8\,\mathcal{R}^{\mu\nu}\, \mathcal{D}_\mu X^I
    \,\mathcal{D}_\nu \bar X^J \nonumber\\
    &\qquad\qquad -\big[\varepsilon^{ik}\, Y_{ij}{}^I\,(F^{+ab\,J}
    -\tfrac14 X^J T^{ab}{}_{lm}\varepsilon^{lm} )\, 
     R(\mathcal{V})_{ab}{}^j{}_k +[\mathrm{h.c.}; I\leftrightarrow J]
     \big]  \Big]\nonumber \\
     &\, +\cdots \,,
\end{align}
where $\mathcal{H}(X,\,\bar X)$ is a real homogeneous function of
degree zero. These invariants, often called `D-terms', are based on a
full superspace integral and they are obviously not encoded in terms
of holomorphic functions. In this particular case the invariant
depends only on the mixed multiple derivatives of $\mathcal{H}(X,\bar
X)$, so that one is dealing with an underlying K\"ahler equivalence,
\begin{equation}
  \label{eq:kahler}
  \mathcal{H}(X,\bar X)\to \mathcal{H}(X,\bar X) +
  \Lambda(X)+\bar\Lambda(\bar X)\,.
\end{equation}
which is based on the fact that a chiral superfield vanishes when
integrated over the full superspace. Obviously the mixed derivative
$\mathcal{H}_{I\bar J}$ can be regarded as a K\"ahler metric. 

It is now straightforward to show that \eqref{eq:real-susp-action}
generates the terms in \eqref{eq:vww3} provided that,
\begin{align}
  \label{eq:Kahler-metric}
  \mathcal{H}_{0\bar{A}}
  =&\, -\tfrac1{384}\mathrm{i}c_A \,|X^0|^{-2} \,,
  \nonumber\\
  \mathcal{H}_{0\bar{0}}=&\,- 
   \tfrac1{384}\mathrm{i}c_A \big(t^A-\bar t^A\big)  \, |X^0|^{-2}\,, 
\end{align}
so that (up to a K\"ahler transformation),
\begin{equation}
  \label{eq:Kahler-pot}
  \mathcal{H}(X,\bar X) = \tfrac1{384} \mathrm{i} c_A \big(t^A\,
  \ln\bar X^0 - \bar t^A \,\ln X^0 \big)\,.  
\end{equation}

\section{The 4D/5D connection and the BPS spinning black hole}
\label{sec:4D-5D-connection-bps-bh}
\setcounter{equation}{0}
In this section we return to some open questions that arose in the
calculation of the entropy of spinning BPS black holes based on $5D$
supersymmetric Lagrangians with higher-derivative couplings. Since the
entropy must be expressed in terms of the charges and the angular
momentum of the black hole, these quantities will have to be
determined as well. The first calculations were carried out in
\cite{Castro:2007hc,Castro:2007ci}, where both the near-horizon
attractor geometry and the field equations leading to the global
solution were studied. The main results were that the reduced $5D$
field equations were inconsistent with the known $4D$ equations, and
that the $5D$ and $4D$ electric charges differ by a constant shift
induced by the higher-order derivative couplings. Another study was
undertaken in \cite{deWit:2009de}. It was aimed at demonstrating that
all the information on charges, angular momentum and entropy can be
obtained from the near-horizon data, and at providing an independent
verification of the results of \cite{Castro:2007hc,Castro:2007ci}. As
it turns out the results of the two studies did not entirely
agree. The precise results for the electric charges were different in
the case of non-zero angular momentum, and furthermore the expressions
for the angular momentum were different. This could have been
interpreted as evidence that the relevant data cannot be determined
from the near-horizon analysis alone. However, this seems unlikely in
view of the fact that the same study in \cite{deWit:2009de} did lead
to a full determination of the entropy, electric charges and angular
momenta for BPS black rings, confirming many independent results based
on field theoretic solutions and on microstate counting
\cite{Gauntlett:2004qy,Bena:2004tk,Cyrier:2004hj,
  Bena:2005ni,Bena:2005ae,Hanaki:2007mb}.

In both of these studies the results were compared to the
corresponding results for four-dimensional black holes, although it is
questionable whether the results should a priori be the same after
straightforward dimensional reduction. Indeed, for black rings it was
noted there are subtle differences between the four- and
five-dimensional charges, and the electric charges are not additively
conserved in five dimensions as a result of the Chern-Simons terms. It
seems likely that the differences between results obtained from
theories that are related by dimensional reduction originate from
topological subtleties related to Chern-Simons terms, which do not
carry over to the lower dimension. Calculations of supergravity
solutions are notoriously difficult in the presence of
higher-derivative interactions, but the results of the present paper
will enable us to confirm once more that this expectation is indeed
correct, as we will demonstrate below in a relatively simple
model. Subsequently we will discuss the topological features
related with the Chern-Simons contributions in more detail.

To examine how the results of these calculations based on the $5D$
supergravity relate to those based on the $4D$ supergravity, we
consider a simple model action,\footnote{
  Unlike in the other sections we do not use Pauli-K\"all\'en
  conventions, but conventions with signature $(-,+,+,+,+)$ and
  $\varepsilon^{01234}=-1$. This leaves our final results
  unchanged} 
\begin{equation}
  \label{eq:model-action}
  8\pi^2 S= \int \mathrm{d}^5x \;\Big\{- E\big(\tfrac12 R +
  \tfrac{1}{4}F_{MN}{} ^2\big )+\frac{\zeta}{128}  \,
  \varepsilon^{MNPQR}\, 
  W_M R_{NP}{}^{AB} R_{QR AB} \Big\}   \,, 
\end{equation}
where $\zeta$ is the strength of the higher-derivative mixed
gauge-gravitational Chern-Simons term and
$F_{MN}=2\, \partial_{[M}W_{N]}$.

Let us assume that this theory has an extremal black hole solution,
whose near-horizon geometry is a fibration of $\mathrm{AdS}_2\times
S^2$, 
\begin{align}
  \label{eq:model-nearhor-extremal} 
  \mathrm{d}s^2=&\, \upsilon_1 \big(-r^2 \mathrm{d}t^2+r^{-2}
  \mathrm{d}r^2 \big)
  +\upsilon_2 \big(\mathrm{d}\theta^2+\sin^2\theta 
  \,\mathrm{d}\varphi^2\big)+\phi^{-2}(\mathrm{d}\psi+B)^2 \,,
  \nonumber \\  
  B=&\, e^0\, r\, \mathrm{d}t + p^0 \cos\theta\
  \mathrm{d}\varphi\,,\nonumber\\ 
  W =&\, W^4 + \chi(\mathrm{d}\psi + B)\,, \nonumber \\
  W^4=&\, e\, r\,\mathrm{d}t + p \cos\theta\,\mathrm{d}\varphi\,, 
\end{align} 
Note that $B$ specifies a value for the Kaluza-Klein gauge field. The
gauge field $W_M$, decomposed according to the standard Kaluza-Klein
ansatz with its fifth component denoted by $\chi$, leads to the field
strengths,
\begin{equation}
  \label{eq:W-F}
  F_{tr}= -(e+\chi\, e^0) \,,\qquad F_{\theta\varphi}=- (p+ \chi\, p^0)
  \sin\theta\,. 
\end{equation}
In what follows we further restrict the background by choosing,
\begin{align}
  \label{eq:backgr-restr}
  p=0\,,\qquad \upsilon_1=\upsilon_2=\upsilon^2\,,\qquad (e^0)^2 +
  (p^0)^2 = \upsilon^2 \phi^2\, .  
\end{align}
In that case we are more in line with the BPS near-horizon horizon
geometry in the full $5D$ supergravity used in \cite{deWit:2009de},
and furthermore the gauge field $W_M$ will be globally defined, which
is important for what follows. With these assumptions the line element
can then be written as,
\begin{align}
  \label{eq:GH-BMPV}
  ds^2=&\,- \phi^{-2}\, \rho^4 \left(p^0\,\mathrm{d}t -
  \frac{e^0}{\rho^2}\Big(\cos\theta\,\mathrm{d}\varphi +
  \frac{1}{p^0}\,\mathrm{d}\psi \Big)\right)^2 \nonumber\\
  &\,
  + \frac{4\,\upsilon^2}{\rho^2}\left(\mathrm{d}\rho^2 +
  \frac{\rho^2}{4}\Big(\mathrm{d}\theta^2  + \mathrm{d}\varphi^2
  +\frac1{(p^0)^2} \,\mathrm{d}\psi^2  + \frac{2}{p^0} 
  \,\cos\theta\,\mathrm{d}\varphi\,\mathrm{d}\psi\Big) \right) \;, 
\end{align}
where we used the definition $\rho= \sqrt{r}$.  To make $p^0$
unambiguous we fix the periodicity interval for $\psi$ to $4\pi$. The
second term of the line element then corresponds to a flat metric, up
to an overall warp factor $4\,\upsilon^2 \rho^{-2}$.  Clearly for $\vert
p^0\vert=1$ we cover the whole four-dimensional space
$\mathbb{R}^4$. For $\vert p^0\vert \not=1$ we have a conical
singularity at the origin. In all cases the three-dimensional horizon
is located at $r=0$ and its cross-sectional area is equal to
\begin{equation}
  \label{eq:area}
  A_3= \int_{\Sigma_\mathrm{hor}} = 16 \pi^2 \upsilon^2 \phi^{-1} \,. 
\end{equation}
The bi-normal $\varepsilon_{MN}$ that characterizes the null surface
at the horizon equals $\varepsilon_{tr} = -\varepsilon_{rt} =
\upsilon^2$.

Subsequently we determine the electric charge associated with the
gauge field $W_M$,
\begin{align}
  \label{eq:5D-charge}
  q^{(5)}=&\, - 4 \big(e +\chi e^0\big) \phi^{-1} -\frac{\zeta}{128\,\pi^2}\,
  \mathcal{Q}_\mathrm{CS} \nonumber\\
  =&\,  - 4 \big(e +\chi e^0\big) \phi^{-1} -\frac{3\,\zeta} {16}\, 
  \frac{p^0 \,(e^0)^2}{[(p^0)^2+(e^0)^2]^2} \,,
\end{align}
where we used the expression for the integrated Chern-Simons term on
the horizon,
\begin{align}
  \label{eq:CS}
  \mathcal{Q}_\mathrm{CS} =&\, \int_{\Sigma_\mathrm{hor}}
   \,
  \mathrm{d}\theta\,\mathrm{d}\varphi\, \mathrm{d}\psi\;
 \,\frac{\varepsilon_{MN}}{2\,v^2} \,
  \varepsilon^{MNPQR}\, \Gamma_{PS}{}^T \big(\partial_Q
  \Gamma_{RT}{}^S
  -\tfrac23 \Gamma_{QT}{}^U \, \Gamma_{RU}{}^S   \big) \nonumber\\
  = &\, 24\pi^2 \, \frac{p^0 \,(e^0)^2}{[(p^0)^2+(e^0)^2]^2} \,.
\end{align}
In what follows this last result will play a crucial role, because the
Chern-Simons term does not transform as a density under
diffeomorphisms. This implies that one may obtain a different answer
upon writing the metric in different coordinates. To ensure that the
coordinate singularity at $r=0$ is not causing complications, we also
evaluated the horizon area \eqref{eq:area} and the Chern-Simons
charge \eqref{eq:CS} in a regular near-horizon metric by converting to
new coordinates, which gave rise to identical results. Nevertheless,
we have also found examples of different metrics which did indeed give
rise to different results for the integrated Chern-Simons term. For
the moment we will proceed, assuming that \eqref{eq:CS} represents the
correct result. At the end of the section we will reconsider this
issue from a topological perspective, which will lend further support to
the correctness of the above result.

We also determine the angular momentum, which, according to
\eqref{eq:model-nearhor-extremal}, will vanish when $e^0=\chi=0$. One
first evaluates the Noether potential for diffeomorphisms parametrized
in terms of a vector $\xi^M$ (the relevant formulae can be obtained
from \cite{deWit:2009de}),
\begin{align}
  \label{eq:N-pot-diff}
  8\pi^2 \,\mathcal{Q}^{MN}(\xi)=&\, \nabla^{[M} \xi^{N]} + F^{MN}\,
  W_P\xi^P  - \tfrac{\zeta}{64}  E^{-1} \big[2\,\varepsilon^{MNPQR}
  W_P\, R_{QR}{}^{ST} \,\nabla_S\xi_T \nonumber \\
  &\,\qquad\qquad - 2\, \varepsilon^{PQRS[M} F_{PQ}
  \, R_{RS}{}^{N]T} \,\xi_T +
  \varepsilon^{PQRST} F_{PQ} \, R_{RS}{}^{MN} \,\xi_T  \Big] \,,
\end{align}
where the Riemann tensor can be written as,
\begin{align}
 \label{eq:Riem-attr}
 \mathcal{R}_{MN}{}^{PQ} =&{}
 -\tfrac12\,\phi^{-2}
 \big(t_{MN}\,t^{PQ} + t_M{}^{[P} \,t_N{}^{Q]} \nonumber\\
&\qquad\qquad
 + 2\,\delta_{[M}{}^{[P} \,t_{N]R}   t^{Q]R}
 -\tfrac12\,\delta_{[M}^P\delta_{N]}^Q \,t_{RT}\,t^{RT}\big)\,,
\end{align} 
with the non-vanishing components of the anti-symmetric tensor $t_{MN}$ given by 
\begin{equation}
  \label{eq:def-t}
  t_{tr}=p^0\,,\qquad t_{\theta\varphi}=-e^0\,\sin{\theta}\,.
\end{equation}
Subsequently one considers the periodic Killing vector
$\partial/\partial\psi$ associated with rotations. In this case we
have $\xi_M \,\mathrm{d}x^M=\phi^{-2}\big(\mathrm{d}\psi +
e^0r\,\mathrm{d}t + p^0\cos\theta\,\mathrm{d}\varphi\big)$, so that
the nonvanishing derivatives of $\xi_M$ equal,  
\begin{equation}
  \label{eq:curl-Killing}
  \nabla_{[t} \xi_{r]} = - \tfrac12 e^0 \,\phi^{-2}\,,\qquad 
  \nabla_{[\theta} \xi_{\varphi]} = - \tfrac12 p^0 \,\phi^{-2}\,. 
\end{equation}
Substituting the above results into \eqref{eq:N-pot-diff} and
integrating over the horizon leads to the following expression for the
angular momentum,
\begin{align}
  \label{eq:angular-mom}
  J_\psi\equiv&\,  \int_{\Sigma_\mathrm{hor}}\,
  \varepsilon_{MN}\,\mathcal{Q}^{MN}(\xi^\psi) \nonumber\\ 
  =&\, -4\,\phi^{-1}\big[\chi e + (\chi^2+ \tfrac12\phi^{-2})
  e^0\big] -\frac{\zeta}{16} \,\frac{p^0[2\, e\,e^0 + 3\chi(3\,(e^0)^2
  -(p^0)^2)]}{[(e^0)^2 +(p^0)^2]^2} \,. 
\end{align}

We would like to briefly compare these results to the results based on
the $4D$ action that one obtains upon dimensional reduction of the
$5D$ action \eqref{eq:model-action}. The corresponding Lagrangian
reads,
\begin{align}
  \label{eq:4D-model-lagr}
  2\pi\,\mathcal{L}=&\, -\sqrt{|g|}\,\phi^{-1}\Big\{\tfrac12 R +
  \tfrac18(\phi^{-2}+2\,\chi^2) F_{\mu\nu}^0{}^2 +\tfrac14
  F_{\mu\nu}{}^2 +\tfrac12\chi\,F_{\mu\nu}^0\,F^{\mu\nu} \Big\}
  \nonumber\\
  &\,+ \tfrac1{128} \zeta\,\varepsilon^{\mu\nu\rho\sigma} \Big\{
  \chi\,R_{\mu\nu}{}^{\lambda\tau} \,R_{\rho\sigma\lambda\tau} +
  \chi\phi^{-2}R_{\mu\nu}{}^{\lambda\tau}\big[ F^0_{\rho\lambda}\,F^0_{\sigma
    \tau}+F^0_{\rho\sigma}\,F^0_{\lambda\tau}\big]
  \nonumber\\
  &\, \quad + \tfrac14 \chi\phi^{-4} F^0_{\mu\nu}\big[2\,F^0_{\rho
    \lambda}\,F^0_{\sigma\tau}
  F^{0\,\lambda\tau} + F^0_{\rho\sigma }\,F^0_{\lambda\tau}{}^2\big] +\tfrac12\chi\phi^{-2}
  \mathcal{D}_\lambda F^0_{\mu\nu} \,\mathcal{D}^\lambda F^0_{\rho\sigma}
   \nonumber\\  
    &\,\quad +  F_{\mu\nu} \Big[
   \phi^{-2} R_{\rho\sigma}{}^{\lambda\tau}\, F^0_{\lambda\tau}
   +\tfrac14\phi^{-4} \big[ F^0_{\lambda\tau}{}^2 
   F^0_{\rho\sigma} + 2\, F^{0\,\lambda\tau}
   F^0_{\rho\lambda}\,F^0_{\sigma\tau}\big] \Big] \Big\} \,,
\end{align}
where $F_{\mu\nu}^0=2\,\partial_{[\mu}B_{\nu]}$.  Eventually we assume
constant values for $v$, $\phi$ and $\chi$, and therefore we have
suppressed above the contributions from space-time derivatives of
these fields. Furthermore, we have absorbed an overall factor of
$4\pi$ to account for the length of the interval of the extra $5D$
coordinate $\psi$.

The $4D$ line element follows from \eqref{eq:GH-BMPV}, 
\begin{equation}
  \label{eq:4d-line-elt}
  \mathrm{d}s^2= \upsilon^2 \big(-r^2 \mathrm{d}t^2+r^{-2}
  \mathrm{d}r^2  + \mathrm{d}\theta^2+\sin^2\theta 
  \,\mathrm{d}\varphi^2\big)\,.
\end{equation}
Therefore the near-horizon geometry equals $\mathrm{AdS}_2\times S^2$
and the Riemann tensor decomposes into the Riemann tensors associated
with each of the two maximally symmetric factors. In addition we have
the field strengths,
\begin{equation}
  \label{eq:D4-field-strengths}
  F^0_{tr} =-e^0\,, \qquad F^0_{\theta\varphi}= - p^0 \sin\theta\,,\qquad
   F_{tr} = -e  \,, \qquad F_{\theta\varphi}= 0 \,. 
\end{equation}
Making use of this fact we determine the value of the $4D$ electric
charges associated with the $4D$ gauge fields $B_\mu$ and $W_\mu$,
\begin{align}
  \label{eq:4D-charges}
  q^{(4)} =&\,-4\phi^{-1} \big(e+\chi e^0\big) - \frac{\zeta}{16} \,
  \frac{p^0\big(5\,(e^0)^2 + 2\,(p^0)^2\big)} {[(e^0)^2 +(p^0)^2]^2}
  \,,  \nonumber\\[.2ex]  
  q_0=&\, -4\,\phi^{-1}\big[\chi e + (\chi^2+ \tfrac12\phi^{-2})
  e^0\big] -\frac{\zeta}{16} \,\frac{p^0[2\, e\,e^0 + 3\chi(3\,(e^0)^2
  -(p^0)^2)]}{[(e^0)^2 +(p^0)^2]^2} \,. 
\end{align}
Comparing these charges to the results for the five-dimensional charge
and angular momentum, specified by \eqref{eq:5D-charge} and
\eqref{eq:angular-mom}, respectively, we find,
\begin{align} 
  \label{eq:charge-diffs}
  q^{(4)}=&\, q^{(5)} - \frac\zeta{8}\,\frac{p^0}{(e^0)^2+ (p^0)^2}\,,
  \nonumber\\ 
  q_0 =&\, J_\psi \,.
\end{align}
The value of the four-dimensional charge $q_0$, associated with the
Kaluza-Klein gauge field, coincides exactly with the five-dimensional
angular momentum of the spinning black hole. On the other hand, the
five-dimensional charge associated with the vector multiplet,
$q^{(5)}$, differs from the four-dimensional charge $q^{(4)}$ which is
obtained after the straightforward reduction of the Lagrangian. And
furthermore this difference is directly related to the Chern-Simons
term. These conclusions are consistent with the results derived in
\cite{deWit:2009de}. In \cite{Castro:2007ci} it was also found that
the five- and four-dimensional charges are not the same and are
related by a shift. However, as it turns out, the latter shift is different
from the one above, because it does not depend on $e^0$. 

In the remainder of this section we explain how this last phenomenon
can be understood in terms of the topology associated with the
Chern-Simons term. The latter arises from the defining condition,
\begin{equation}
  \label{RR}
  \mathrm{d} \,\mathrm{Tr}\,\big[C \wedge \mathrm{d}C- \tfrac23 C\wedge C\wedge
  C\big] =\tfrac14 \mathrm{Tr}\,\big[R\wedge R\big]\,, 
\end{equation}
where $C$ is an appropriate matrix-valued connection equal to the
Christoffel or to the spin connection. If there is a non-trivial flux
of $R\wedge R$ along a 4-cycle, it is not possible to define the
Chern-Simons term globally, but only on patches connected by an
appropriate closed, non-exact, gauge transformation. Here we may
regard the Chern-Simons term as a composite 3-form potential
constructed from the metric, rather than as a fundamental gauge
field. The gauge transformation between the patches is thus induced by
certain diffeomorphisms, where we assume that any additional ambiguity
associated with the underlying diffeomorphisms will have no
cohomological consequences.

In four dimensions, the flux of $R\wedge R$ is a topological
invariant, so that the Chern-Simons term cannot be defined as a gauge
potential, as it would carry no degrees of freedom. However, when the
Chern-Simons term is viewed as a 3-form gauge potential in five
space-time dimensions, one finds that its magnetic dual is a scalar
field. The situation is thus analogous to that of a magnetic monopole
located at some point in a three-dimensional Euclidean space, except
that here we are dealing with a point in a five-dimensional
space-time. Likewise, from that point in space-time there will be a
Dirac string emanating from it, extending to infinity. Obviously
this extension to infinity has to be compatible with the fact that we
are dealing with stationary solutions. We recall that the choice
for the string is associated with a certain regular gauge patch and
the transition functions between different patches are provided by
gauge transformations (in this case induced by diffeomorphisms). The
strings associated with two different patches are connected by a
corresponding two-dimensional surface which encodes the corresponding
gauge transformation.

We are interested in computing the integral of the Chern-Simons term
in \eqref{RR} over a spatial 3-dimensional surface, as a contribution
to the electric charge. In the analogous situation of the monopole in
three spatial dimensions, this integral is a Wilson line whose value
may differ on each patch depending on whether it encircles the Dirac
string or not. The difference is given by the gauge transformation
between the patches and is proportional to the flux of the
corresponding field strength given by $R\wedge R$.

Without loss of generality, we may employ two patches and impose that
the corresponding Dirac strings are timelike and intersect at most
once with each spatial slice.\footnote{ 
  This condition can be relaxed by introducing more than two
  patches.} 
The question is then how precisely the string will extend through
space-time. In the context of a black hole background, there are only
two acceptable choices, namely, that the string will move to spatial
infinity at large time, or that the string will remain behind the
black hole horizon, so that the unphysical string singularity is not
observable.

It follows that the surface connecting the strings associated with two
different sections define a semi-infinite plane along the time and
radial directions. The closed but non-exact gauge transformation,
$\beta^0$, which connects the two patches, is given by the normal form
$\beta^0$ of this plane
\cite{Henneaux:1986tt,Bekaert:2002cz,Bekaert:2002eq}. For the metric
above, this reads
\begin{equation}
  \label{eq:concecting-gauge-tr}
    \beta^0=\frac{p^0}{(e^0)^2+(p^0)^2}\sin\theta\, \mathrm{d}\theta \wedge
    \mathrm{d} \varphi\wedge \,\mathrm{d}\psi\,.
\end{equation}
Here, the factor $p^0$ is implied by the induced metric on the $S^3$
defined by constant values of $t$ and $r$, and the overall
normalisation is fixed by demanding that in the static limit the
integral over $\beta^0$ is equal to $2/p^0$, i.e.~equal to the flux of
$\mathrm{Tr}\,[R\wedge R]$ for a Gibbons-Hawking base space. Therefore,
the value of the Chern-Simons term in the patch that contains the
Dirac string singularity, is related to the value in the regular patch,
where the singularity is located at infinity, by 
\begin{align}
  \label{gau-shift}
  \mathrm{Tr}\, \big[ C_{[\mu}\partial_\nu C_{\rho]} - \tfrac23
  C_{[\mu}\, C_{\nu}\,C_{\rho]} \big]\Big|_{\text{sing}}= &\,
  \mathrm{Tr}\,\big[C_{[\mu}\partial_\nu C_{\rho]} - \tfrac23 C_{[\mu}\,
  C_{\nu}\,C_{\rho]} \big]\Big|_{\text{reg}}+ \beta^0_{\mu\nu\rho}\,.
\end{align}
The gauge transformation in \eqref{gau-shift} changes the position of the
intersection of the Dirac string within a given spatial slice. We
conclude that the integral of the above Chern-Simons term over a
3-surface can take two distinct values depending whether the
intersection point is contained or not in that 3-surface. Therefore,
the electric charges of BPS spinning black holes in a five-dimensional
theory containing a mixed gauge/gravitational Chern-Simons term can
also take two different values, depending on how the patches are chosen.

In an asymptotically flat five-dimensional setting one can impose
regularity in the bulk of the solution, pushing the Dirac brane to
infinity. The connection used to evaluate the integrated Chern-Simons
term in \eqref{eq:CS} in consistent with this requirement, in line
with the general view that a non-trivial Taub-NUT charge is not
considered to be part of the black hole in the center, so that no
singularity associated to $R\wedge R$ should appear. This parallels
the approach used in the microscopic counting, where the large-charge
limit is taken for the electric charges but not for the Taub-NUT
charge.

One can now also consider the corresponding four-dimensional solution
based on \eqref{eq:4D-model-lagr}. In that setting, it is not
acceptable to have a gauge-dependent singularity present near spatial
infinity. However, a physical solution can still be obtained if the
singularity is hidden behind the horizon, which amounts to a change of
patch as in \eqref{gau-shift}. From a four-dimensional perspective,
this corresponds to the addition of a delta source singularity
interpreted as a magnetic monopole associated with the Kaluza-Klein
gauge field.

The results above are in agreement with the results from the
near-horizon analysis for spinning five-dimensional BPS black holes
given in \cite{deWit:2009de}, where it was found that the difference
between the four- and five-dimensional charges differ by a shift that
depends both on $p^0$ and on $e^0$. In that case the angular momentum
is proportional to $e^0$, so that the shift will depend on the angular
momentum. This result differs from that in \cite{Castro:2007ci}, in
spite of the fact that there the singularity associated with the
Taub-NUT charge has also been moved to spatial infinity. However, for
reasons that are not clear to us, the coefficient of that singularity
depended only on $p^0$, so that the difference between the four- and
five-dimensional electric charges was constant and did not depend on
the angular momentum.

\section{Concluding remarks}
\label{sec:conclusions}
\setcounter{equation}{0}
In this paper we studied the off-shell dimensional reduction of
five-dimensional $N=1$ conformal supergravity to four space-time
dimensions. We obtained the full dictionary expressing the $5D$
fields in terms of the $4D$ fields and showed in some detail how to
connect to the standard $N=2$ superconformal Lagrangians in four
dimensions. The advantage of performing the reduction off-shell is
that it allows one to make a precise comparison of the two theories
beyond the usual two-derivative actions, by connecting the
four-derivative invariants on both sides.

Somewhat unexpectedly, we found that upon reduction of the
four-derivative supersymmetric action in five dimensions, which contains
terms quadratic in the Riemann tensor, one finds terms that are not
compatible with the two four-derivative $N=2$ supersymmetric
invariants known in four dimensions. In this way we deduce the
presence of at least one new four-dimensional invariant that is quadratic in
the Ricci tensor, whose complete structure remains to be
uncovered. Terms like these will be required when considering the
$N=2$ supersymmetric extension of the Gauss--Bonnet invariant.

As a further application of our reduction scheme, we studied the
effect of the mixed gau\-ge/gra\-vi\-ta\-tio\-nal Chern-Simons term,
contained in the five-dimensional four-derivative action, on the
definition of the electric charge for spinning supersymmetric black
holes. Consistent with previous results on five-dimensional BPS black
holes \cite{deWit:2009de}, we find a shift in the charges upon
dimensional reduction to four dimensions, whose value depends on the
angular momentum. Similar examples of this phenomenon have been
discovered at the two-derivative level for black rings. Just as in
that case, the subtleties can be understood from the presence of Dirac
branes and the associated sections.

The off-shell approach to dimensional reduction developed in this
paper is not specific to five dimensions and can be applied to other
situations. One interesting example would be the reduction from four
to three dimensions, especially in connection to the c-map.

\section*{Acknowledgement}
We thank Daniel Butter, Michael Duff, Marc Grisaru, Finn Larsen and
Ashoke Sen for valuable discussions. N.B. acknowledges the hospitality
of the Indian Institute of Science (Bangalore) and the Tata Institute
of Fundamental Research (Mumbai) extended to her during the completion
of this work. N.B. is supported by a Veni grant of the `Nederlandse
Organisatie voor Wetenschappelijk Onderzoek (NWO)'. B.d.W. is
supported by the ERC Advanced Grant no. 246974, {\it ``Supersymmetry:
  a window to non-perturbative physics''}. The work of S.K. is
supported by the French ANR contract 05-BLAN-NT09-573739, the ERC
Advanced Grant no.  226371, the ITN programme PITN-GA-2009-237920 and
the IFCPAR programme 4104-2. Until October 1 2011 it was part of the
research programme of the `Stichting voor Fundamenteel Onderzoek der
Materie (FOM)', which is financially supported by the `Nederlandse
Organisatie voor Wetenschappelijk Onderzoek (NWO)'.

\begin{appendix}
%
\section{Relations between 5D and 4D Riemann curvatures}
\label{App:5-4D-Riemann-curv}
\setcounter{equation}{0}
Based on \eqref{eq:kk-ansatz} one can evaluate the relation between
$5D$ and $4D$ curvature components. In the equations below,
derivatives $\mathcal{D}_a$ are covariant with respect to $4D$ local
Lorentz transformations and dilatations. The results are as follows
(in this appendix the $5D$ curvature components are consistently
denoted by $\hat R$),
\begin{align}
  \label{eq:curvatures-world}
  \hat R_{\mu\nu}{}^{ab} =& \, R_{\mu\nu}{}^{ab} +\tfrac12 \phi^{-2} \Big[
  F(B)_{\mu}{}^{[a} \,F(B)_\nu{}^{b]} + F(B)_{\mu\nu} F(B)^{ab}
  \Big]\nonumber\\
  &\, - B_{[\mu} \Big[2\, \phi^{-3} F(B)_{\nu]}{}^{[a}
  \, \mathcal{D}^{b]} \phi + \mathcal{D}_{\nu]}[\phi^{-2} F(B)^{ab}]
  \Big]\,, \nonumber\\[.4ex]
  \hat R_{\mu\nu}{}^{a5} =&\,
  - \mathcal{D}_{[\mu}[\phi^{-1} F(B)_{\nu]}{}^a] -
  \phi^{-2} \mathcal{D}^a\phi\, F(B)_{\mu\nu}\nonumber\\
  &\,
  +B_{[\mu}\Big[2\,\mathcal{D}_{\nu]} [\phi^{-2}\,\mathcal{D}^a\phi] + \tfrac12
  \phi^{-3}\,F(B)_{\nu]b} \, F(B)^{ab}\Big]  \,,  \nonumber\\[.4ex]
  \hat R_{\mu\hat 5}{}^{ab} =&\,\tfrac12 \mathcal{D}_\mu [\phi^{-2}
  F(B)^{ab} ] + \phi^{-3} F(B)_\mu{}^{[a} \,\mathcal{D}^{b]}\phi\,,
  \nonumber\\[.4ex]
  \hat R_{\mu\hat 5}{}^{a5} =&\,- \mathcal{D}_\mu
  [\phi^{-2} \mathcal{D}^a\phi] - \tfrac14 \phi^{-3} \,F(B)_{\mu
    b}F(B)^{ab}  \,.
\end{align}
With tangent-space indices, $\hat R_{CD}{}^{AB}$ takes the form,
\begin{align}
  \label{eq:curvatures-tangent}
  \hat R_{cd}{}^{ab} =& \, R_{cd}{}^{ab} +\tfrac12 \phi^{-2} \Big[
  F(B)_{c}{}^{[a} \,F(B)_d{}^{b]} + F(B)_{cd} F(B)^{ab}
  \Big]\,, \nonumber\\[.4ex]
  \hat R_{cd}{}^{a5} =&\,
  \tfrac12 \phi^{-1} \,\mathcal{D}^aF(B)_{cd}  -
  \phi^{-2} \Big[\mathcal{D}^a\phi\, F(B)_{cd} -  
  \,F(B)^a{}_{[c} \mathcal{D}_{d]}\phi \Big]   \,,  \nonumber\\[.4ex]
  \hat R_{c 5}{}^{ab} =&\, 
  \tfrac12\phi^{-1} \, \mathcal{D}_c F(B)^{ab} -
  \phi^{-2}\Big[F(B)^{ab}\, \mathcal{D}_{c}\phi - F(B)_c{}^{[a}
  \,\mathcal{D}^{b]}\phi \Big]\,,   \nonumber\\[.4ex]
  \hat R_{c5}{}^{a5} =&\,-\phi\, D_c(\omega)
  [\phi^{-2} \mathcal{D}^a\phi] - \tfrac14 \phi^{-2} \,F(B)_{cb}F(B)^{ab}
  \,.
\end{align}
Note that these components satisfy the pair-exchange property of the
Riemann tensor.  Contracted versions of the Riemann tensor take the
form,
\begin{align}
  \label{eq:contracted-R}
  \hat R_{cB}{}^{aB} =&\, R_{cb}{}^{ab} + \tfrac12 \phi^{-2}
  F(B)_{cb}F(B)^{ab} -\phi\, \mathcal{D}_c [\phi^{-2}\mathcal{D}^a\phi] \,,
  \nonumber\\
  \hat R_{A5}{}^{Ab} =&\, \tfrac12\phi^{-1} \, \mathcal{D}_a F(B)^{ab}
  -\tfrac32 \phi^{-2}\,F(B)^{ab}\, \mathcal{D}_{a}\phi\,,\nonumber\\
    \hat R_{A5}{}^{A5}=&\, -\phi\, \mathcal{D}_a(\omega)
  [\phi^{-2} \mathcal{D}^a\phi] - \tfrac14 \phi^{-2}
  \,F(B)_{ab}F(B)^{ab}\,, \nonumber\\
  \hat R_{AB}{}^{AB} =&\, R_{ab}{}^{ab} - 2\, \phi\,
  \mathcal{D}_a [\phi^{-2} \mathcal{D}^a\phi]
  +\tfrac14 \phi^{-2} \,F(B)_{ab} F(B)^{ab}  \,.
\end{align}
Furthermore one may consider components of $\hat R_{[AB}{}^{EF} \,\hat
R_{CD]EF}$, which are required for the dimensional reduction of the
$5D$ mixed Chern-Simons term, 
\begin{align}
  \label{eq:RwedgeR}
  \hat R_{[ab}{}^{EF}\,\hat R_{cd]EF} =&\, R_{[ab}{}^{ef} \,R_{cd]ef}
  +
  \phi^{-2}R_{ab}{}^{ef}\big[F(B)_{ce}F(B)_{df}+F(B)_{cd}F(B)_{ef}\big]
  \nonumber\\
  &\, + \tfrac14 \phi^{-4} F(B)_{ab}\big[2\,F(B)_{ce}F(B)_{df}
  F(B)^{ef} +
  F(B)_{cd}F(B)^2\big] \nonumber\\
  &\,
  +\tfrac12\phi^{-2} \mathcal{D}_eF(B)_{ab} \,\mathcal{D}^eF(B)_{cd} \nonumber\\
  &\, +2 \phi^{-1} \mathcal{D}^e F(B)_{ab} \big[F(B)_{ce}\,
  \mathcal{D}_d\phi^{-1} +
  F(B)_{cd} \,\mathcal{D}_e\phi^{-1}\big] \nonumber\\
  &\, + 2\, F_{ab}(B)\,\big[F_{cd}(B) \,(\mathcal{D}_e\phi^{-1} )^2
  +2\, F(B)_{ce}\,\mathcal{D}^e\phi^{-1}
  \,\mathcal{D}_d\phi^{-1}\big]\, \Big\vert_{[abcd]} \,,
  \nonumber\\[.4ex]
   \hat R_{\hat 5[a}{}^{EF}\,\hat R_{cd]EF} =&\,- \phi\,\mathcal{D}_{[a}\Big[
   \tfrac12\phi^{-2} R_{cd]}{}^{ef} F(B)_{ef} \nonumber\\
   &\qquad +\tfrac18\phi^{-4} \big[ F(B)^2
   F(B)_{cd]} + 2\, F(B)^{ef} F(B)_{ce}F(B)_{df}\big]  \nonumber\\
   &\qquad -2\, \phi^{-1}F(B)_{c}{}^{e}
   \,\mathcal{D}_{d]}(\mathcal{D}_e\phi^{-1}) 
   + F(B)_{cd]}  (\mathcal{D}\phi^{-1})^2 \Big] \,.  
\end{align}
where we made use of the Bianchi identity on $F(B)$ on the $4D$ Riemann
tensor.

\section{Conversion of $5D$ symplectic Majorana spinors into $4D$
  chiral spinors}
\label{sec:spinors-four-five}
\setcounter{equation}{0}
In this paper we have to convert $5D$ symplectic Majorana spinors into
$4D$ chiral spinors, so as to obtain the dimensionally reduced
supersymmetry transformations in $4D$ notation. We explain this here
using Pauli-K\"all\'en notation where all gamma matrices are
hermitian. As it turns out, one can use the same $4\times4$
charge-conjugation matrix $C$ in four and in five dimensions,
satisfying
\begin{equation}
  \label{eq:charge-conj-matrix}
  C^\mathrm{T}=-C\,,\qquad C^\dagger= C^{-1}\,. 
\end{equation}
The $4D$ gamma matrices are subject to, 
\begin{align}
  \label{eq:4D-clifford}
  C\gamma_aC^{-1}=&\, -\gamma_a{}^\mathrm{T}\,, \qquad \gamma_5 =
  \tfrac1{24} \varepsilon^{abcd}\gamma_a \gamma_b \gamma_c \gamma_d \,,
\end{align}
where the indices $a,b,\ldots$ take four values. As is obvious,
$\gamma_5$ satisfies $C\gamma_5C^{-1} =
\gamma_5{}^\mathrm{T}$. Majorana spinors $\psi$ have chiral components
satisfying, 
\begin{equation}
  \label{eq:4majorana}
  C^{-1} \big(\bar\psi_\mp\big)^\mathrm{T}  = \psi_\pm  \,.  
\end{equation}
where $\psi_\pm$ are eigenspinors of $\gamma_5$ with eigenvalue $\pm
1$. Since we will always be dealing with R-symmetry doublets of
spinors labeled by an index $i,j,\ldots=1,2$, one uses the convention
that the position of the index denotes at the same time the
chirality. For reasons of convenience this is not done in a uniform
way, so that the relation between the chirality and the position of
the index differs from spinor to spinor.

In five dimensions, the charge conjugation properties of the gamma
matrices $\hat\gamma_A$ are,
\begin{align}
  \label{eq:5D-clifford}
  C\hat \gamma_AC^{-1}=&\, \hat\gamma_A{}^\mathrm{T}\,, \qquad
  \hat\gamma_{ABCDE} 
  = \mathbf{1} \,\varepsilon_{ABCDE} \,,
\end{align}
where the indices $A,B,\ldots$ take five values. Note that the last
equation defines $\gamma_5$ as a product of the remaining four gamma
matrices. However, these gamma matrices are not identical to the $4D$
ones, in view of the sign difference between the first equations of
\eqref{eq:4D-clifford} and \eqref{eq:5D-clifford}. Nevertheless, as we
shall see below, the matrix $\gamma_5$ will remain the same as the
$4D$ one, defined in the second equation of \eqref{eq:4D-clifford}.

It is easy to construct the remaining $5D$ gamma matrices from the
$4D$ ones. Namely, one may assume that,
\begin{equation}
  \label{eq:54gamma}
  \hat\gamma_a = \mathrm{i}\gamma_a\gamma_5\,, 
\end{equation}
since the matrices on the right-hand side are symmetric with respect
to charge conjugation, they anti-commute with $\gamma_5$, and the
product of two gamma matrices satisfies, $\hat\gamma_a\hat\gamma_b=
\gamma_a\gamma_b$. Because the gamma matrices are different, also the
definition of the Dirac conjugate will differ for $4D$ and $5D$
spinors, according to the relation, 
\begin{equation}
  \label{eq:45D-conjugate}
  \bar\psi\big|_{D=5} =  \bar\psi\big|_{D=4} \, \mathrm{i}\gamma_5\,.  
\end{equation}
This fact will be relevant for the action when reducing to four
dimensions, but also when relating the $5D$ symplectic Majorana
condition to the $4D$ Majorana condition. 

The $5D$ symplectic Majorana condition for spinors $\psi^i$ read, 
\begin{equation}
  \label{eq:sym-maj}
  C^{-1} \big(\bar\psi_{i} \big)^\mathrm{T}  = \varepsilon_{ij}\,
  \psi^j \,.   
\end{equation}
Upon replacing the $5D$ Dirac conjugate to the $4D$ one, one obtains, 
\begin{equation}
  \label{eq:4D-sym-maj}
  C^{-1} \big(\bar\psi_{i\mp} \big)^\mathrm{T}  =\mp\mathrm{i} 
  \varepsilon_{ij}\, \psi^j_\pm  \,, 
\end{equation}
where we have adopted chiral spinor components. Now let us assume that
the $4D$ spinor with upper index $i$ is of{\it positive} chirality, so
that we identify it with the $5D$ field $\psi^i_+$. In the $4D$
context we know from \eqref{eq:4majorana}, that the Dirac conjugate is
then equal to the corresponding field of negative chirality, which we
write with a lower $\mathrm{SU}(2)$ index. In this way we derive from
\eqref{eq:4D-sym-maj} that the $5D$ field can be decomposed in $4D$
chiral spinors according to
\begin{equation}
  \label{eq:upper-pos-chir}
  \psi^i\big|_{5D} = \psi^i_+ + \mathrm{i} \varepsilon^{ij} \psi_{j-}\,,
\end{equation}
where $\psi^i$ and $\psi_i$ appearing on the right-hand side are the
positive- and negative-chirality components of a $4D$ Majorana spinor
doublet. 

In case the four-dimensional {\it negative-chirality} spinor carries
an upper $\mathrm{SU}(2)$ index, then the above relation changes into, 
\begin{equation}
  \label{eq:upper-neg-chir}
  \psi^i\big|_{5D} = \psi^i_- - \mathrm{i} \varepsilon^{ij} \psi_{j+}\,. 
\end{equation}

After this conversion defined by \eqref{eq:upper-pos-chir} and \eqref{eq:upper-neg-chir} it remains possible to redefine the $4D$
Majorana spinors by a chiral $\mathrm{U}(1)$ and a scale
transformation without affecting the Majorana condition.

\section{Supersymmetry transformations in four dimension}
\label{App:SC}
\setcounter{equation}{0}
In four space-time dimensions we follow the notation used {\it e.g.}
in \cite{LopesCardoso:2000qm,deWit:2010za}. Space-time and Lorentz
indices are denoted by $\mu,\nu,\ldots$, and $a,b,\ldots$,
respectively; $\mathrm{SU}(2)$-indices are denoted by $i,j,\ldots$.
Furthermore, (anti-)sym\-metrizations are always defined with unit
strength. 

For the convenience of the reader we summarize the $4D$ transformation
rules of the superconformal fields and their relation to the
superconformal algebra, as well as their covariant quantities
contained in the so-called Weyl supermultiplet. The superconformal
algebra comprises the generators of the general-coordinate, local
Lorentz, dilatation, special conformal, chiral $\mathrm{SU}(2)$ and
$\mathrm{U}(1)$, supersymmetry (Q) and special supersymmetry (S)
transformations.  The gauge fields associated with general-coordinate
transformations ($e_\mu{}^a$), dilatations ($b_\mu$), chiral symmetry
($\mathcal{V}_\mu{}^i{}_j$ and $A_\mu$) and Q-supersymmetry
($\psi_\mu{}^i$) are independent fields.  The remaining gauge fields
associated with the Lorentz ($\omega_\mu{}^{ab}$), special conformal
($f_\mu{}^a$) and S-supersymmetry transformations ($\phi_\mu{}^i$) are
dependent fields.  They are composite objects, which depend on the
independent fields of the multiplet \cite{deWit:1980tn}. The
corresponding supercovariant curvatures and covariant fields are
contained in a tensor chiral multiplet, which comprises $24+24$
off-shell degrees of freedom. In addition to the independent
superconformal gauge fields, it contains three other fields: a
Majorana spinor doublet $\chi^i$, a scalar $D$, and a selfdual Lorentz
tensor $T_{abij}$, which is anti-symmetric in $[ab]$ and $[ij]$. The
Weyl and chiral weights have been collected in table \ref{table:weyl}.
%
\begin{table}[t]
\begin{tabular*}{\textwidth}{@{\extracolsep{\fill}}
    |c||cccccccc|ccc||ccc| }
\hline
 & &\multicolumn{9}{c}{Weyl multiplet} & &
 \multicolumn{2}{c}{parameters} & \\[1mm]  \hline \hline
 field & $e_\mu{}^{a}$ & $\psi_\mu{}^i$ & $b_\mu$ & $A_\mu$ &
 $\mathcal{V}_\mu{}^i{}_j$ & $T_{ab}{}^{ij} $ &
 $ \chi^i $ & $D$ & $\omega_\mu^{ab}$ & $f_\mu{}^a$ & $\phi_\mu{}^i$ &
 $\epsilon^i$ & $\eta^i$
 & \\[.5mm] \hline
$w$  & $-1$ & $-\tfrac12 $ & 0 &  0 & 0 & 1 & $\tfrac{3}{2}$ & 2 & 0 &
1 & $\tfrac12 $ & $ -\tfrac12 $  & $ \tfrac12  $ & \\[.5mm] \hline
$c$  & $0$ & $-\tfrac12 $ & 0 &  0 & 0 & $-1$ & $-\tfrac{1}{2}$ & 0 &
0 & 0 & $-\tfrac12 $ & $ -\tfrac12 $  & $ -\tfrac12  $ & \\[.5mm] \hline
 $\gamma_5$   &  & + &   &    &   &   & + &  &  &  & $-$ & $ + $  & $
 -  $ & \\ \hline
\end{tabular*}
\vskip 2mm
\renewcommand{\baselinestretch}{1}
\parbox[c]{\textwidth}{\caption{\label{table:weyl}{\footnotesize
Weyl and chiral weights ($w$ and $c$) and fermion
chirality $(\gamma_5)$ of the Weyl multiplet component fields and the
supersymmetry transformation parameters.}}}
\end{table}

Under Q-supersymmetry, S-supersymmetry and special conformal
transformations the independent fields of the Weyl multiplet transform
as follows,
\begin{align}
  \label{eq:weyl-multiplet}
  \delta e_\mu{}^a  =&\, \bar{\epsilon}^i \, \gamma^a \psi_{ \mu i} +
  \bar{\epsilon}_i \, \gamma^a \psi_{ \mu}{}^i \, , \nonumber\\
  \delta \psi_{\mu}{}^{i} =&\, 2 \,\mathcal{D}_\mu \epsilon^i - \tfrac{1}{8}
  T_{ab}{}^{ij} \gamma^{ab}\gamma_\mu \epsilon_j - \gamma_\mu \eta^i
  \, \nonumber \\
  \delta b_\mu =&\, \tfrac{1}{2} \bar{\epsilon}^i \phi_{\mu i} -
  \tfrac{3}{4} \bar{\epsilon}^i \gamma_\mu \chi_i - \tfrac{1}{2}
  \bar{\eta}^i \psi_{\mu i} + \mbox{h.c.} + \Lambda^a_K e_{\mu a} \, ,
  \nonumber \\
  \delta A_{\mu} =&\, \tfrac{1}{2} \mathrm{i} \bar{\epsilon}^i \phi_{\mu i} +
  \tfrac{3}{4} \mathrm{i} \bar{\epsilon}^i \gamma_\mu \, \chi_i +
  \tfrac{1}{2} \mathrm{i}
  \bar{\eta}^i \psi_{\mu i} + \mbox{h.c.} \, , \nonumber\\
  \delta \mathcal{V}_\mu{}^{i}{}_j =&\, 2\, \bar{\epsilon}_j
  \phi_\mu{}^i - 3
  \bar{\epsilon}_j \gamma_\mu \, \chi^i + 2 \bar{\eta}_j \, \psi_{\mu}{}^i
  - (\mbox{h.c. ; traceless}) \, , \nonumber \\
  \delta T_{ab}{}^{ij} =&\, 8 \,\bar{\epsilon}^{[i} R(Q)_{ab}{}^{j]} \,
  , \nonumber \\
  \delta \chi^i =&\, - \tfrac{1}{12} \gamma^{ab} \, \Slash{D} T_{ab}{}^{ij}
  \, \epsilon_j + \tfrac{1}{6} R(\mathcal{V})_{\mu\nu}{}^i{}_j
  \gamma^{\mu\nu} \epsilon^j -
  \tfrac{1}{3} \mathrm{i} R_{\mu\nu}(A) \gamma^{\mu\nu} \epsilon^i + D
  \epsilon^i +
  \tfrac{1}{12} \gamma_{ab} T^{ab ij} \eta_j \, , \nonumber \\
  \delta D =&\, \bar{\epsilon}^i \,  \Slash{D} \chi_i +
  \bar{\epsilon}_i \,\Slash{D}\chi^i \, .
\end{align}
Here $\epsilon^i$ and $\epsilon_i$ denote the spinorial parameters of
Q-supersymmetry, $\eta^i$ and $\eta_i$ those of S-supersymmetry, and
$\Lambda_K{}^a$ is the transformation parameter for special conformal
boosts.  The full superconformally covariant derivative is denoted by
$D_\mu$, while $\mathcal{D}_\mu$ denotes a covariant derivative with
respect to Lorentz, dilatation, chiral $\mathrm{U}(1)$, and
$\mathrm{SU}(2)$ transformations,
\begin{equation}
  \label{eq:D-epslon}
  \mathcal{D}_{\mu} \epsilon^i = \big(\partial_\mu - \tfrac{1}{4}
    \omega_\mu{}^{cd} \, \gamma_{cd} + \tfrac1{2} \, b_\mu +
    \tfrac{1}{2}\mathrm{i} \, A_\mu  \big) \epsilon^i + \tfrac1{2} \,
  \mathcal{V}_{\mu}{}^i{}_j \, \epsilon^j  \,.
\end{equation}

Just as in five dimensions the gauge fields associated with local
Lorentz transformations, S-supersymmetry and special conformal boosts,
$\omega_{\mu}{}^{ab}$, $\phi_\mu{}^i$ and $f_{\mu}{}^a$, respectively,
are composite and determined by conventional constraints. In this case
these constraints are S-supersymmetry invariant and they take the
following form,
\begin{align}
  \label{eq:conv-constraints}
  &R(P)_{\mu \nu}{}^a =  0 \, , \nonumber \\[1mm]
  &\gamma^\mu R(Q)_{\mu \nu}{}^i + \tfrac32 \gamma_{\nu}
  \chi^i = 0 \, , \nonumber\\[1mm]
  &
  e^{\nu}{}_b \,R(M)_{\mu \nu a}{}^b - \mathrm{i} \tilde{R}(A)_{\mu a} +
  \tfrac1{8} T_{abij} T_\mu{}^{bij} -\tfrac{3}{2} D \,e_{\mu a} = 0
  \,.
\end{align}
The curvatures appearing in \eqref{eq:conv-constraints} take the
following form, 
\begin{align}
  \label{eq:curvatures-4}
  R(P)_{\mu \nu}{}^a  = & \, 2 \, \partial_{[\mu} \, e_{\nu]}{}^a + 2 \,
  b_{[\mu} \, e_{\nu]}{}^a -2 \, \omega_{[\mu}{}^{ab} \, e_{\nu]b} -
  \tfrac1{2} ( \bar\psi_{[\mu}{}^i \gamma^a \psi_{\nu]i} +
  \mbox{h.c.} ) \, , \nonumber\\[.2ex]
  R(Q)_{\mu \nu}{}^i = & \, 2 \, \mathcal{D}_{[\mu} \psi_{\nu]}{}^i -
  \gamma_{[\mu}   \phi_{\nu]}{}^i - \tfrac{1}{8} \, T^{abij} \,
  \gamma_{ab} \, \gamma_{[\mu} \psi_{\nu]j} \, , \nonumber\\[.2ex]
  R(M)_{\mu \nu}{}^{ab} = & \,
  \, 2 \,\partial_{[\mu} \omega_{\nu]}{}^{ab} - 2\, \omega_{[\mu}{}^{ac}
  \omega_{\nu]c}{}^b
  - 4 f_{[\mu}{}^{[a} e_{\nu]}{}^{b]}
  + \tfrac12 (\bar{\psi}_{[\mu}{}^i \, \gamma^{ab} \,
  \phi_{\nu]i} + \mbox{h.c.} ) \nonumber\\
& \, + ( \tfrac14 \bar{\psi}_{\mu}{}^i   \,
  \psi_{\nu}{}^j  \, T^{ab}{}_{ij}
  - \tfrac{3}{4} \bar{\psi}_{[\mu}{}^i \, \gamma_{\nu]} \, \gamma^{ab}
  \chi_i
  - \bar{\psi}_{[\mu}{}^i \, \gamma_{\nu]} \,R(Q)^{ab}{}_i
  + \mbox{h.c.} ) \,. 
\end{align}

Chiral multiplets can be consistently reduced by imposing a reality
constraint, which requires specific values for the Weyl and
chiral weights. The two cases that are relevant are the vector
multiplet, which arises upon reduction from a scalar chiral multiplet,
and the Weyl multiplet, which is a reduced anti-selfdual chiral tensor
multiplet. Both reduced multiplets require weight $w=1$.

The vector multiplet contains a complex scalar $X$, a chiral spinor
$\Omega_i$ and a gauge field $W_\mu$, which transform under Q- and
S-supersymmetry transformations as follows,
\begin{align}
  \label{eq:variations-vect-mult}
  \delta X =&\, \bar{\epsilon}^i\Omega_i \,,\nonumber\\
  \delta\Omega_i =&\, 2 \Slash{D} X\epsilon_i
     +\ft12 \varepsilon_{ij}  F_{\mu\nu}^-
   \gamma^{\mu\nu}\epsilon^j +Y_{ij} \epsilon^j
     +2X\eta_i\,,\nonumber\\
  \delta W_{\mu} = &\, \varepsilon^{ij} \bar{\epsilon}_i
  (\gamma_{\mu} \Omega_j+2\,\psi_{\mu j} X)
  + \varepsilon_{ij}
  \bar{\epsilon}^i (\gamma_{\mu} \Omega^{j} +2\,\psi_\mu{}^j
  \bar X)\,,\nonumber\\
\delta Y_{ij}  = &\, 2\, \bar{\epsilon}_{(i}
  \Slash{D}\Omega_{j)} + 2\, \varepsilon_{ik}
  \varepsilon_{jl}\, \bar{\epsilon}^{(k} \Slash{D}\Omega^{l)
  } \,,
\end{align}
Here $F_{\mu\nu}^-$ denotes the anti-selfdual component associated with
the field strength of $W_\mu$, 
\begin{align}
  \label{eq:vector-fs}
    F_{ab}^- =&   \big(\delta_{ab}{}^{cd} -\tfrac12
    \varepsilon_{ab}{}^{cd}\big) e_c{}^\mu e_d{}^\nu \,\partial_{[\mu}
    W_{\nu]}\nonumber\\
    &\,
  +\tfrac14\big[\bar{\psi}_{\rho}{}^i\gamma_{ab} \gamma^\rho\Omega^{j}
  + \bar{X}\,\bar{\psi}_\rho{}^i\gamma^{\rho\sigma}\gamma_{ab}
  \psi_\sigma{}^j
  - \bar{X}\, T_{ab}{}^{ij}\big]\varepsilon_{ij}  \,, 
\end{align}

The Weyl multiplet was already discussed at the beginning of this
appendix. The Weyl and chiral weights for the vector multiplet, the
hypermultiplet and the Weyl multiplet in four dimensions has been
summarized in tables \ref{table:w-weights-matter-4D} and
\ref{table:weyl}.

\end{appendix}

\providecommand{\href}[2]{#2}

\end{document}